\documentclass[superscriptaddress, onecolumn, amsmath, amssymb, 11pt, aps]{revtex4-2}

\usepackage{graphicx}
\usepackage{dcolumn}
\usepackage{bm}
\usepackage{amsmath}
\usepackage[colorlinks,linkcolor=blue,anchorcolor=blue,citecolor=blue]{hyperref}
\usepackage{color}

\usepackage{epstopdf}
\usepackage{float}

\begin{document}

\title{Acoustic Pancharatnam-Berry Geometric Phase}

\author{Wanyue Xiao}
\affiliation{Department of Physics, City University of Hong Kong, Kowloon, Hong Kong, China}

\author{Wenjian Kuang}
\affiliation{Division of Science, Engineering and Health Studies, College of Professional and Continuing Education, Hong Kong Polytechnic University, Hong Kong, China}

\author{Sibo Huang}
\affiliation{Department of Electrical Engineering, City University of Hong Kong, Kowloon, Hong Kong, China}

\author{Shanjun Liang}
\email{junot.liang@cpce-polyu.edu.hk} 
\affiliation{Division of Science, Engineering and Health Studies, College of Professional and Continuing Education, Hong Kong Polytechnic University, Hong Kong, China}

\author{Din Ping Tsai}
\affiliation{Department of Electrical Engineering, City University of Hong Kong, Kowloon, Hong Kong, China}

\author{Shubo Wang} 
\email{shubwang@cityu.edu.hk} 
\affiliation{Department of Physics, City University of Hong Kong, Kowloon, Hong Kong, China}

\begin{abstract}
\textbf{Geometric phases provide a unified framework for understanding diverse phenomena in quantum and classical physics. The Pancharatnam-Berry (PB) geometric phase, arising from variation of optical transverse polarization, has transformed light manipulation. However, this phase has never been observed in sound waves due to their curl-free longitudinal nature. Here, we theoretically and experimentally demonstrate that the PB phase can emerge in general inhomogeneous sound waves with polarization evolution of velocity field. Using surface sound waves as an example, we uncover the intriguing Janus property of the PB phase arising from spin-momentum locking, and realize acoustic PB metasurfaces for versatile wavefront manipulation. We further extend the mechanism to free-space structured sound and realize acoustic $q$-plate for generating acoustic vortices through spin-orbit interaction. Our work provides new insights into sound wave properties and enables the manipulation of inhomogeneous acoustic fields via the PB phase, with potential applications in acoustic communications and imaging.}
\end{abstract}

\maketitle

\noindent\textbf{\large{Introduction}}\\
\noindent Geometric phase arises when the eigenstate of a system undergoes evolutions in the parameter space \cite{1berry1984quantal}. It had a profound impact on physics with an elegant interpretation based on the fiber bundle theory \cite{2gonoskov2022charged,58cohen2019geometric}. The geometric phase can reveal intricate topological structures of the state and parameter spaces \cite{3PhysRevLett.51.2167} and give rise to intriguing phenomena such as the Aharonov–Bohm effect \cite{4PhysRev.115.485}, quantum Hall and quantum spin-Hall effects 
\cite{5PhysRevLett.49.405,6PhysRevLett.95.226801}. Recently, the geometric phases in classical wave systems have attracted enormous attention. A prominent example is the geometric phase \textcolor{black}{ induced by Bloch state evolution in the momentum space of periodic optical and acoustic systems}. This type of geometric phase characterizes the topological properties of nontrivial edge or corner states \cite{9RevModPhys.91.015006,10xue2022topological}, which have applications in communications \cite{11wang2009observation,12yang2020terahertz}, lasing \cite{13bandres2018topological,14yang2022topological}, and quantum information processing \cite{15mittal2018topological,16barik2018topological}.

In addition to the momentum-space geometric phase, there is another type of geometric phase \textcolor{black}{that emerges in real space}---the PB phase \cite{21pancharatnam1956generalized,22berry1987adiabatic},  due to the evolution of optical transverse polarization induced by anisotropic materials or structures \cite{31bomzon2002space,32marrucci2006optical,56Xiao:24}. This geometric phase plays a pivotal role in light manipulation by optical metasurfaces, which has remarkable applications such as metalens \cite{33wang2018broadband,34chen2018broadband}, holographic imaging \cite{35zheng2015metasurface, li2018addressable}, and nonlinear harmonic generations \cite{36li2015continuous}. Despite extensive research on the PB phase, this phase has not been observed in sound waves (in air or fluids), which are curl-free longitudinal waves lacking \textcolor{black}{transverse} polarization degrees of freedom. \textcolor{black}{Notably, the evolution of acoustic orbital angular momentum (OAM), achieved by modulating the \textit{global} vortex pattern with complex structures, can give rise to a real-space geometric phase \cite{43liu2021acoustic,44liu2022experimental,45zhang2023geometric,chen2024super}. However, this phase is not the PB phase, which originates from the \textit{local} polarization property. }

\textcolor{black}{Sound waves} comprise a scalar pressure field $p$ and a vector velocity field $\mathbf{v}$. The velocity field can induce intriguing acoustic phenomena in a way similar to electromagnetic fields \cite{47shi2019observation,46PhysRevB.99.174310,64long2020symmetry,48wang2021spin,69PhysRevLett.129.204301,64PhysRevLett.131.114001,70tong2023topological}. An interesting  question is: Can sound waves carry the PB phase through the velocity field? An affirmative answer seems counter-intuitive, considering the different nature of light and sound. Light is a transverse wave with two vector-field degrees of freedom (i.e., electric field $\mathbf{E}$ and magnetic field $\mathbf{H}$). Light-matter and sound-matter interactions involve fundamentally different physics. Specifically, the optical PB phase appears in circularly polarized light interacting with anisotropic structures such as metasurfaces, where the subwavelength meta-atoms induce polarization evolution through the electric dipole. In contrast, general acoustic velocity fields are not circularly polarized globally, and their interaction with subwavelength meta-atoms is usually dominated by the acoustic monopole, which has isotropic mode fields and cannot induce polarization evolution. 

In this article, we theoretically and experimentally demonstrate that the PB geometric phase can emerge in inhomogeneous sound waves due to the polarization evolution of velocity field. Using \textcolor{black}{surface sound waves (SSWs)} as an example, we show that the interaction between \textcolor{black}{circularly polarized velocity fields}  and meta-atoms with \textcolor{black}{dominant} dipole response can induce the PB phase covering 2$\pi$ full range. The acoustic PB phase exhibits a Janus property originating from the spin-momentum locking—it has different values for the SSWs propagating in opposite directions. \textcolor{black}{Leveraging this geometric phase, we design and realize acoustic PB metasurfaces for nearly arbitrary wavefront manipulation. The mechanism can be readily extended to other inhomogeneous acoustic fields in free space, including Bessel beams. We show that the PB phase can be employed to realize acoustic $q$-plates, which can enable intriguing acoustic spin-orbit interaction and induce the conversion of vortex topological charge in the Bessel beams.}\\

\noindent\textbf{\large{Inhomogeneous sound waves}}\\
\noindent \noindent One common type of inhomogeneous sound waves is the SSWs, which can appear on the surface of structured substrates \textcolor{black}{\cite{zhu2011holey, 50PhysRevApplied.11.034061}}. We consider a rigid lossless substrate immersed in air, which has square holes of depth $l$ and side length $w$ forming a two-dimensional square lattice in the $xy$ plane with period $q$, as depicted in Fig. 1A. The substrate supports SSWs propagating in the $xy$ plane. \textcolor{black}{We focus on the deep subwavelength regime with $\lambda \gg q$. Under this condition,} the SSWs have an isotropic dispersion $\beta=k_0 \sqrt{1+\left(w/q\right)^4 \tan ^2\left(k_0 l\right)}$ \textcolor{black}{below the free-space sound line $k_0=\omega/c$}, where $\beta$ is the propagation constant \textcolor{black}{(see supplementary text)}. This analytical dispersion relation is shown in Fig. 1B as the solid red line, which agrees with the simulation result (symbols). For the SSW propagating in $\pm x$ direction, the velocity field is $\mathbf{v}=\left(v_{x},~0, ~v_{z}\right)$ = $\left(\pm n_{\mathrm{eff}},~0,~i \gamma\right) e^{ \pm i k_{0} n_{\mathrm{eff}} x-k_{0} \gamma z}$, where $n_{\text {eff }}=\beta / k_0$ is the effective index, and $\gamma=\sqrt{n_{\text {eff}}^{2}-1}$ characterizes the field's decay. Since $v_{x}$ and $v_{z}$ have a $\pi/2$ phase difference, $\mathbf{v}$ is elliptically polarized, as depicted in Fig. 1C for the SSW propagating in $+x$ direction. Thus, the SSW carries a \textcolor{black}{transverse spin in $-y$ direction \cite{47shi2019observation,64long2020symmetry}}. By the time-reversal symmetry, the SSW propagating in $-x$ direction carries a transverse spin in $+y$ direction. This locking between the spin direction and propagation direction is known as the spin-momentum locking or transverse spin-orbit interaction \cite{52bliokh2015transverse,55van2016universal,swang19}. \textcolor{black}{Without loss of generality, we choose the working frequency marked by the dashed line in Fig. 1B. The mechanism also applies to other frequencies.}

\begin{figure}[htp] 
	\centering
\includegraphics[width=0.7\linewidth]{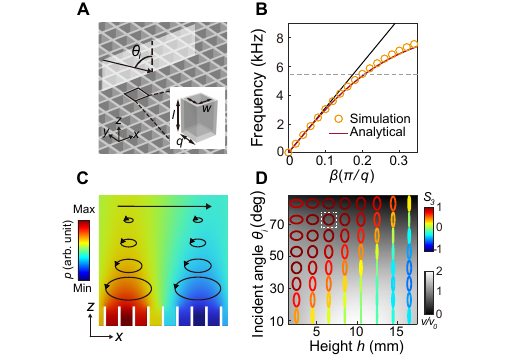} 

	\caption{\textcolor{black}{\textbf{Inhomogeneous sound waves.}} ($\mathbf{A}$) The substrate drilled with periodic square blind holes. The inset shows the unit cell. ($\mathbf{B}$) Dispersion relation of the SSWs supported by the holey substrate. 
    The black solid line denotes the sound dispersion in free space. 
    ($\mathbf{C}$) Velocity polarization ellipses of the SSW \textcolor{black}{at the frequency corresponding to the dashed line in $\mathbf{B}$}. ($\mathbf{D}$) Polarization and normalized amplitude of the background velocity field at different $h$ (the height above the substrate) and incident angles $\theta_{i}$\textcolor{black}{, due to the interference of incident and reflected fields}.}
	\label{Fig. 1} 
\end{figure}

\textcolor{black}{Inhomogeneous sound waves also appear in the interference of freely propagating waves \cite{47shi2019observation}.} We consider a plane wave illuminates the holey substrate with incident angle $\theta_{i}$, as shown in Fig. 1A. The incident wave will interfere with the reflected wave, giving rise to an inhomogeneous total velocity field $\mathbf{v}$ above the substrate. Figure 1D shows the numerically determined polarization ellipses of $\mathbf{v}$ as a function of $h$ (the height above the substrate) and $\theta_{i}$. The color of the ellipses denotes the Stokes parameter $S_{3}$ characterizing the \textcolor{black}{ellipticity} \cite{71jackson2012classical}. The background color shows $v / v_{0}$ with $v=|\mathbf{v}|$ and $v_{0}$ being the incident amplitude. Clearly, the polarization and amplitude of $\mathbf{v}$ depend on both $h$ and $\theta_{i}$. For a fixed $\theta_{i}$, $S_{3}$ varies with $h$ and can change sign across the interference pattern; \textcolor{black}{$S_{3}=+1$ (corresponding to circularly polarized velocity field) at certain $h$ (marked by the white dashed box).} Therefore, both the SSWs and the interference wave \textcolor{black}{exhibit elliptically or circularly polarized velocity fields, which can give rise to the PB phase as a result of polarization evolution.}\\

\begin{figure}[!t] 
	\centering
\includegraphics[width=0.7\linewidth]{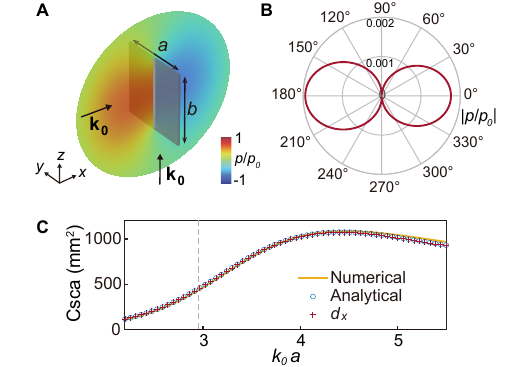} 

	\caption{\textcolor{black}{\textbf{Scattering properties of the meta-atom with \textcolor{black}{dominant} dipole response.}} ($\mathbf{A}$) Scattered pressure near field of the meta-atom plate. Two plane waves with $\pi / 2$ phase difference incident along $x$ and $z$ directions. ($\mathbf{B}$) Scattered pressure amplitude in the far field. ($\mathbf{C}$) Scattering cross section of the meta-atom. The dashed line marks the frequency for the results in $\mathbf{A}$ and $\mathbf{B}$.}
	\label{Fig. 2} 
\end{figure}

\noindent\textbf{\large{Meta-atoms with \textcolor{black}{dominant} dipole response}}

\noindent The polarization evolution of velocity field can be induced by acoustic dipole meta-atoms. However, most subwavelength meta-atoms have a dominant monopole mode with isotropic mode field that cannot induce velocity polarization evolution (see fig. S4). Therefore, it is necessary to design acoustic meta-atoms supporting \textcolor{black}{dominant} dipole mode. We consider the thin rigid plate in Fig. 2A under the incidence of the velocity field $\mathbf{v}=$ $\frac{p_{0}}{\sqrt{2} \rho_{0} c}\left(e^{i k_{0} x}, 0, i e^{i k_{0} z}\right)$. Figure 2A also shows the normalized scattered pressure $p/p_{0}$ on the $x z$ plane in the near field. Figure 2B shows $|p/p_{0}|$ in the far field. As noticed, both the near-field and far-field pressure characteristics correspond to a \textcolor{black}{dominant} dipole mode. The scattering cross section contributed by the dipole moment $\mathbf{d}$ can be analytically determined as $C_{\mathrm{sca}}^{\mathbf{d}}=\frac{k_{0}^{2}}{24 \pi \rho_{0} c I_{0}}|\mathbf{d}|^{2}$, where $I_{0}$ is the incident intensity, $\mathbf{d}=\oint \mathbf{n}p_t \mathrm{d} A$ with $\mathbf{n}$ being the surface unit normal vector and $p_{t}$ being the surface total pressure (see supplementary text). As shown in Fig. 2C, the analytical result agrees well with the numerically simulated total scattering cross section of the plate, confirming the \textcolor{black}{dominant dipole response of the meta-atom plate in a broad band of frequencies}. The cross symbols denote the contribution of the dipole component $d_{x}$ to the scattering cross section, indicating that the dipole is perpendicular to the plate. Therefore, the thin plate can serve as the meta-atom to manipulate the velocity field polarization through its dipole mode.\\

\noindent\textbf{\large{Acoustic PB geometric phase}}\\
\noindent We construct a metasurface by arranging the meta-atoms on the holey substrate periodically along $y$ direction with a period of $p$, as shown in Fig. 3A. The orientation of the meta-atoms in $x z$ plane is characterized by the angle $\alpha$ with respect to $x$ axis. A plane wave obliquely incidents in $x z$ plane with angle $\theta_{i}$. As discussed earlier, the interference of the incident and reflected fields gives rise to a circularly polarized velocity field at the positions of the meta-atoms, \textcolor{black}{which carries a transverse spin in $-y$ direction (denoted by the white circle with an arrow)}. This background velocity field excites the linearly polarized dipole $\mathbf{d}$ in each meta-atom. The dipole field couples to the SSWs propagating in $+x$ and $-x$ directions, denoted as $+$SSW and $-$SSW, respectively. This process is accompanied by the variations of the velocity polarization, which induces the PB geometric phase. 

\begin{figure}[t!] 
\centering
\includegraphics[width=\linewidth]{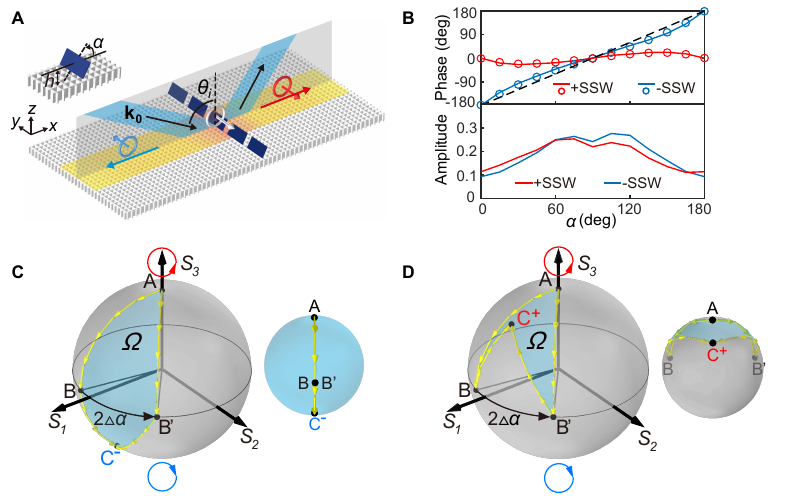} 

\caption{\textcolor{black}{\textbf{Acoustic PB geometric phase and Poincaré sphere interpretation.}} ($\mathbf{A}$) Acoustic PB metasurface under the incidence of a plane wave with incident angle $\theta_{i}$. The inset shows the meta-atom. ($\mathbf{B}$) The PB phases and amplitudes of the SSWs for different rotation angles $\alpha$. The solid lines denote the simulation results. The circles denote the results obtained by evaluating the solid angle in the Poincaré sphere. Evolution trajectories of velocity polarization on the Poincaré sphere for the SSWs propagating in ($\mathbf{C}$) $-x$ and ($\mathbf{D}$) $+x$ directions. \textcolor{black}{The
smaller spheres show the maximum solid angle that can be achieved in each case.}}
	\label{Fig. 3} 
\end{figure}

We determine the PB phase for different rotation angles $\alpha$ by numerically simulating the SSWs' phase. The results are shown in the upper panel of Fig. 3B, \textcolor{black}{where the red solid line denotes the PB phase $\Phi_{\text{PB}}^{+}$ for $+$SSW and the blue solid line denotes the PB phase $\Phi_{\text{PB}}^{-}$ for $-$SSW}. Remarkably, the PB phase exhibits a Janus property: $\Phi_{\text{PB}}^{+}$ and $\Phi_{\text{PB}}^{-}$ have different values. $\Phi_{\text{PB}}^{-}$ can cover $2 \pi$, but $\Phi_{\text{PB}}^{+}$ cannot. This is attributed to the spin flipping that happens when the background field is converted to $-$SSW. In contrast, no spin flipping happens to +SSW. The situation will be reversed if the spin of the background field is reversed, which can be realized by simply changing the incident angle to $-\theta_i$. We notice that $\Phi_{\text{PB}}^{-}$ slightly deviates from the linear relation $\Phi_{\text{PB}}^{-}=2 \alpha$ (dashed line) due to the elliptical polarization of the SSWs. The elliptical polarization also results in different amplitudes of the SSWs at different $\alpha$, as shown in the lower panel of Fig. 3B. \textcolor{black}{Uniform wave amplitudes can be achieved by tailoring the geometry of the meta-atoms or the substrate holes (see figs. S5 and S6)}.\\

\noindent\textbf{\large{Poincaré sphere interpretation}}\\
\noindent The acoustic PB phase can be intuitively understood with the Poincaré sphere describing the polarization of vector fields. \textcolor{black}{Figure 3 (C and D) shows the polarization evolutions of the velocity field for $-$SSW and +SSW, respectively.} The polarization of the background velocity field corresponds to the north pole (point A). The velocity polarization of $\pm$SSW can be characterized by the Stokes vector $\mathbf{S}^{\mathrm{s}}=\left(\frac{1}{n_{\text {eff }}^{2}+\gamma^{2}}, 0, \frac{ \pm 2 \gamma n_{\mathrm{eff}}}{n_{\text {eff }}^{2}+\gamma^{2}}\right)$, corresponding to the point $\text{C}^{\pm}$. The acoustic dipole $\mathbf{d}$ of the meta-atom has $\mathbf{S}^{\mathrm{d}}=[\cos (2 \alpha), \sin (2 \alpha), 0]$, corresponding to the point B on the equator. A variation of the meta-atom's rotation angle $(\Delta \alpha)$ changes its polarization from $\mathrm{B}$ to $\mathrm{B}^{\prime}$. The corresponding change of the PB phase equals half of the solid angle $\Omega$ subtended by the area enclosed by the loop A $\rightarrow \mathrm{B} \rightarrow \mathrm{C}^{\pm}\rightarrow \mathrm{B}^{\prime} \rightarrow \mathrm{A}$ \cite{21pancharatnam1956generalized,22berry1987adiabatic}. 
We evaluate $\Omega$ to obtain the geometric phases of $-$SSW and $+$SSW. The results are denoted by the blue and red circles in Fig. 3B, respectively, which are consistent with the numerical results (solid blue and red lines) obtained by directly simulating the SSWs' phase. \textcolor{black}{The smaller spheres in Fig. 3 (C and D) show the maximum solid angle that can be achieved in each case. The maximum solid angle for $-\mathrm{SSW}$ is $4 \pi$ because $\mathrm{C}^-$ locates on the lower half sphere. In contrast, the maximum solid angle cannot reach $4 \pi$ for $+\mathrm{SSW}$ because $\mathrm{C}^+$ locates on the upper half sphere.}\\

\noindent\textbf{\large{Wavefront manipulation and experiments}}\\
\noindent The acoustic PB phase \textcolor{black}{can serve as a convenient} mechanism for nearly arbitrary manipulation of the SSWs' wavefront. As shown in Fig. 4 (A and B), we design acoustic PB metasurfaces to demonstrate two typical wavefront manipulations, i.e., anomalous deflection and focusing, for $-$SSW since $\Phi_{\text{PB}}^{-}$ can cover $2\pi$. We conduct both full-wave simulations and experiments. The experimental set-up is shown in Fig. 4C. The substrate and metasurface are fabricated by 3D printing. A speaker array is used to generate the incident plane wave, and a microphone mounted on a moving stage is used to measure the pressure field of the SSW. 

\begin{figure}[!t] 
	\centering
\includegraphics[width=\linewidth]{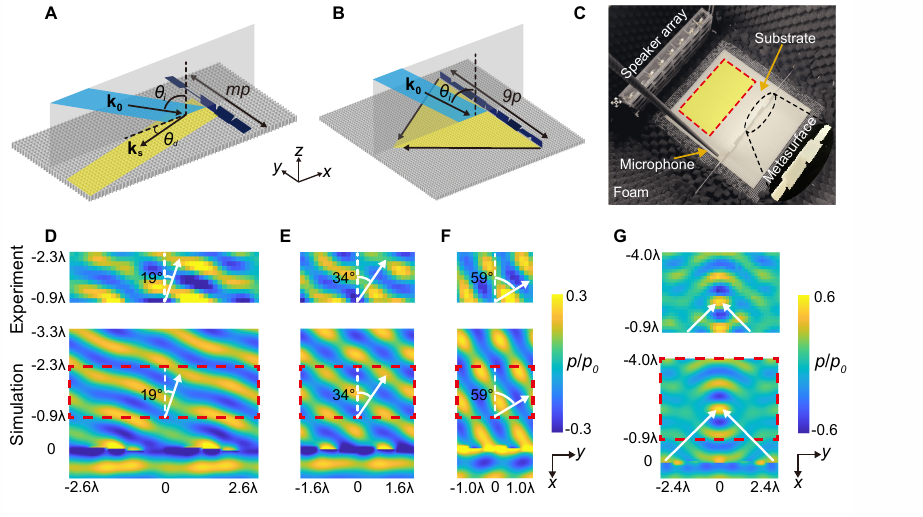} 

	\caption{\textcolor{black}{\textbf{Wavefront manipulation by the acoustic PB metasurfaces.}} ($\mathbf{A}$) Steering and ($\mathbf{B}$) focusing of the SSW by acoustic PB metasurfaces. ($\mathbf{C}$) Experimental set-up. The measurement area is yellow colored. ($\mathbf{D}$ to $\mathbf{F}$) Experimentally measured (upper panels) and simulated (lower panels) pressure fields of the SSW deflected by three metasurfaces with different phase gradients. Each metasurface has a supercell comprising $m$ meta-atoms. ($\mathbf{G}$) Experimentally measured (upper panel) and simulated (lower panel) pressure fields of the focused SSW.}
	\label{Fig. 4} 
\end{figure}

For the anomalous deflection in Fig. 4A, we design three PB metasurfaces with different phase gradients $k_{g}=\Delta \Phi_{\text{PB}}^{-} / p$. Here, $\Delta \Phi_{\text{PB}}^{-}$ is PB phase difference between two nearby meta-atoms. The SSW will be deflected by angle $\theta_{d}=$ $\sin ^{-1}\left(k_{g} / \beta\right)$, according to the generalized Snell's law \cite{60yuscience,assouar2018acoustic}. The upper panels of Fig. 4 (D to F) shows the experimentally measured pressure field of the SSW, which agree well with the simulation results in the lower panels (i.e., regions enclosed by the red rectangles). The analytical results of $\theta_{d}$ are indicated by the white arrows, which agree with the simulated and experimentally measured wavefronts. 

For the focusing in Fig. 4B, we design a PB metalens with geometric phase profile $\Phi_{\text{PB}}^{-}(y)=-\beta\left(\sqrt{y^2-f^2}-f\right)$ \cite{59PhysRevApplied.2.064002}, where $f$ is the focal length and $y$ denotes the location of the meta-atoms. The metalens can convert the incident plane wave to the SSW converging at a desired focal point. Figure 4G shows the experimental (upper panel) and simulation (lower panel) results for the SSW's pressure field, which agree well with each other. We observe the focusing of the SSW with a focal length $f=2.1 \lambda$. The analytically predicted focal length based on $\Phi_{\text{PB}}^{-}(y)$ is indicated by the white arrows, which is consistent with the experimental and simulation results. These results demonstrate the capability of the acoustic PB phase in manipulating the SSWs. \\

\noindent\textbf{\large{Acoustic $q$-plate and spin-orbit interaction}}\\
\noindent The acoustic PB phase is not limited to surface waves but generally exists in inhomogeneous acoustic fields. Here, we apply the mechanism to free-space acoustic Bessel beams. We consider the Bessel beam with pressure field $p=A J_{l}(\kappa r) e^{i l \varphi+i k_{z} z}$, where $A$ is the amplitude, $k_{z}=k_{0} \cos \theta_{0}$ is the longitudinal wavevector, $\kappa=$ $k_{0} \sin \theta_{0}$ is the transverse wavevector, $\theta_{0}$ is aperture angle, and $l$ denotes the topological charge (i.e., OAM quantum number) of the acoustic vortex associated with the Bessel beam \cite{46PhysRevB.99.174310}. \textcolor{black}{The beam has the transverse velocity field}
\begin{equation}
\mathbf{v}=B_{l}(\hat{\mathbf{x}}+i \hat{\mathbf{y}}) e^{i(l-1) \varphi+i k_{z} z}+C_{l}(\hat{\mathbf{x}}-i \hat{\mathbf{y}}) e^{i(l+1) \varphi+i k_{z} z},
\end{equation}
\noindent where $B_{l}=-i A^{\prime} \kappa J_{l-1}(\kappa r)$ and $C_{l}=i A^{\prime} \kappa J_{l+1}(\kappa r)$ with $A^{\prime}=A /(\rho_{0} \omega)$. The velocity field has two components carrying opposite spin $\sigma= \pm 1$ and different OAM $m=l \mp 1$. Using $|\sigma, m\rangle$ to denote the two components, the velocity field can be rewritten as:
\begin{equation}
\mathbf{v}=B_{l}|1, l-1\rangle+C_{l}|-1, l+1\rangle.    
\end{equation}
\noindent \textcolor{black}{Equation (2) indicates that the velocity \textcolor{black}{polarization} is decided by the relative magnitude of $B_{l}$ and $C_{l}$, which depends on the radial distance $r$. Thus, the \textcolor{black}{polarization} varies in space, and its distribution has a cylindrical symmetry.}

\begin{figure}[!t] 
	\centering
\includegraphics[width=0.58\linewidth]{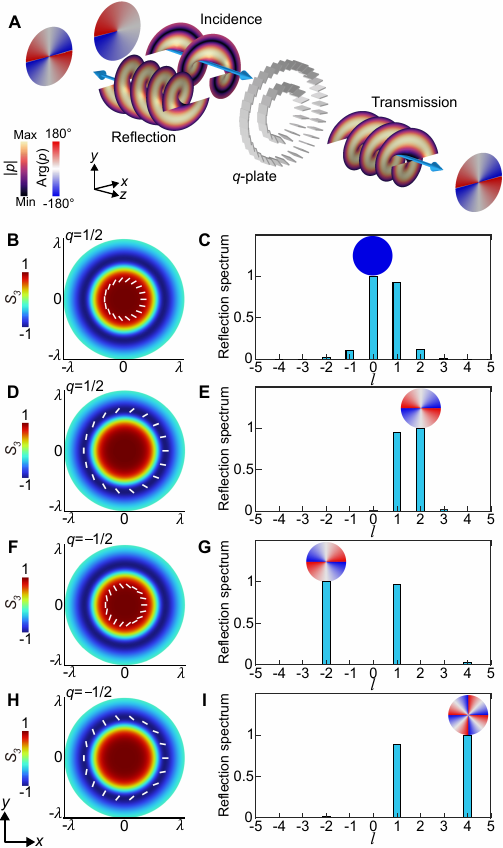} 

	\caption{\textcolor{black}{\textbf{Vortex topological charge conversion by the acoustic $q$-plates.}}  ($\mathbf{A}$) The phase of the pressure field for the incident Bessel beam with \textcolor{black}{topological charge} $l=$ 1. Interaction between the incident Bessel beam and acoustic $q$-plates with different topological charges: ($\mathbf{B}$),($\mathbf{D}$) $q=1 / 2$ and ($\mathbf{F}$),($\mathbf{H}$) $q=$ $-1 / 2$. The colormaps display the Stokes parameter $S_{3}$ for the \textcolor{black}{transverse velocity field of the} incident beam. ($\mathbf{C}$), ($\mathbf{E}$), ($\mathbf{G}$), ($\mathbf{I}$) Reflection spectra in the angular-momentum Fourier space. The insets show the phase distribution of the pressure field for the new components \textcolor{black}{generated by the $q$-plates}.}
	\label{Fig. 5} 
\end{figure}

\textcolor{black}{We consider the incident Bessel beam interacts with an acoustic PB metasurface, as shown in Fig. 5A. The metasurface comprises a circular array of the meta-atom plates, each rotated by an angle $\alpha(\varphi)=q \varphi$ with respect
to $x$ axis, where $q$ is the topological charge of the metasufaces and $\varphi$ denotes the azimuthal coordinate of the meta-atom plates. \textcolor{black}{This metasurface ``$q$-plate"} generates scattering fields in both forward and backward directions, corresponding to the transmission and reflection. The fields can exhibit a topological charge different from that of the incident Bessel beam, due to the spin-orbit interaction enabled by the $q$-plate. We will focus on the reflected field to demonstrate the effect.}

\textcolor{black}{Without loss of generality, we set $l=1$ for the incident Bessel beam. In this case, the pressure field $p=A J_{1}(\kappa r) e^{i \varphi+i k_{z} z}$ exhibits a phase vortex in $x y$-plane, as shown in Fig. 5A. The velocity field $\mathbf{v}=B_{1}|1,0\rangle+C_{1}|-1,2\rangle$ exhibits inhomogeneous \textcolor{black}{polarization} characterized by the Stokes parameter $S_{3}$, as shown by the color in Fig. 5B. We notice that $S_{3} \approx$ 1 in the center region because the local velocity field is dominated by the left-handed circularly polarized (LCP) component $|1,0\rangle$, and $S_{3} \approx-1$ in the outer region because the local velocity field is dominated by the right-handed circularly polarized (RCP) component $|-1,2\rangle$.} 

\textcolor{black}{
Figure 5B shows the $q$-plate with $q=1 / 2$ locating in the center region of the Bessel. Figure 5C shows the reflection spectrum in the angular-momentum Fourier space. We notice that a new component with $l=0$ is generated in the reflected field, besides the original component of $l=1$. The phase distribution of the pressure field for this new component is shown in the inset of Fig. 5C. Then, we consider the $q$-plate with $q=1 / 2$ positioned in the outer region of the Bessel beam, as shown in Fig. 5D. In this case, the reflected field contains a new component with $l=2$, as shown in Fig. 5E, where the inset shows the phase distribution of the pressure field for this new component. We also construct the $q$-plate with topological charge $q=-1 / 2$ and place it in the center and outer regions, as shown in Fig.5 (F and H), respectively. In the former case, the $q$-plate generates a new component of $l=-2$ in the reflected field, as shown in Fig. 5G. In the latter case, the $q$-plate generates a new component of $l=4$ in the reflected field, as shown in Fig. 5I. These results demonstrate that the acoustic $q$-plates can generate reflected fields with different OAM depending on the velocity spin and the topological charge $q$ \textcolor{black}{of the $q$-plates}. Similar phenomena also exist in the transmitted fields.}

\textcolor{black}{
The above phenomena are attributed to the PB phase \textcolor{black}{induced by velocity polarization evolution}. When the $q$-plate locates in the center region with \textcolor{black}{dominating} LCP velocity field $|1,0\rangle$, the reflected field comprises both LCP and RCP velocity field components. \textcolor{black}{The LCP component does not involve spin flipping and does not carry a PB phase. Thus, it has the same $l$ as the incident Bessel beam.} The RCP component involves spin flipping and is imparted with a PB phase. For the $q$-plate in Fig. 5 (B and F), the PB phase is $\Phi_{\mathrm{PB}}=2 \alpha=2 q \varphi$, which introduces additional OAM of $2 q$ to the reflected field. Thus, the reflected field contains a new component $|-1,0+2 q\rangle$. In the case of Fig. 5B with $q=1 / 2$, this new component is $|-1,1\rangle$, corresponding to $l=0$. In the case of Fig. 5F with $q=-1 / 2$, this new component is $|-1,-1\rangle$, corresponding to $l=-2$. When the $q$-plate locates in the outer region, where the local velocity field is \textcolor{black}{dominated by the RCP component} $|-1,2\rangle$, the PB phase becomes $\Phi_{\mathrm{PB}}=-2 \alpha=$ $-2 q \varphi$, which introduces additional OAM of $-2 q$ to the reflected field. Thus, the reflected field contains a new component $|1,2-2 q\rangle$. In the case of Fig. 5D with $q=1 / 2$, this new component is $|1,1\rangle$, corresponding to $l=2$. In the case of Fig. 5H with $q=-1 / 2$, this new component is $|1,3\rangle$, corresponding to $l=4$.}

More generally, for the incident Bessel beam with local velocity field $|\sigma, l-\sigma\rangle$, the $q$-plate converts the spin $\sigma$ to $-\sigma$, while simultaneously imparting \textcolor{black}{$2 \sigma q$ to the OAM of the field}. This gives rise to a new component $\mid-\sigma, l-\sigma+$ $2 \sigma q\rangle$ in the reflected field. \textcolor{black}{We note that the vortex topological charge conversion here is induced by the azimuthal gradient of the PB phase. The OAM-mediated geometric phase also involves vortex topological charge conversion \cite{43liu2021acoustic,44liu2022experimental,45zhang2023geometric,chen2024super}. That conversion, however, is induced by the azimuthal gradient of propagation or resonant phase. We also note that the PB geometric phase is a local phase at each spatial point of polarization evolution, while the OAM-mediated geometric phase is a global phase of the vortex beam. }\\

\noindent\textbf{\large{Discussion}}\\
In conclusion, we demonstrate the acoustic PB phase induced by the velocity \textcolor{black}{polarization evolution} in inhomogenous sound waves, providing the missing piece of acoustic geometric phases. The phase can arise in SSWs interacting with anisotropic meta-atoms, which enables the realization of acoustic PB metasurfaces for nearly arbitrary wavefront manipulation. The mechanism also applies to free-space inhomogeneous acoustic fields, where the PB phase enables the realization of acoustic metasurface $q$-plate. The $q$-plate can induce intriguing acoustic spin-orbit interaction and enable the conversion of vortex topological charge in Bessel beams. Akin to the optical PB phase, the acoustic PB phase is a broadband phenomenon and can also emerge at other frequencies as long as the meta-atoms exhibit a dominant dipole response. The acoustic PB \textcolor{black}{phase} can enable flexible control of inhomogeneous sound waves by simple structures and are particularly suitable for manipulating structured acoustic fields, with potential applications in acoustic communications and imaging.\\

\noindent\textbf{\large{Materials and Methods}}\\
\noindent\textbf{Numerical simulation}\\
The full acoustic wave simulations are performed with the package COMSOL Multiphysics.  For the simulations in Fig.1, we set $w=4.5 \mathrm{~mm}, l=7.5 \mathrm{~mm}$ and $q=5.5 \mathrm{~mm}$ for the holey substrate. For the simulations in Fig. 2, the meta-atom plate has dimensions $a=b=30 \mathrm{~mm}$, and $t=1 \mathrm{~mm}$. For the simulations of the PB metasurface in Fig. 3, we set $a=31 \mathrm{~mm}, b=$ $12.5 \mathrm{~mm}$, $t=1 \mathrm{~mm}$, $h=6.5 \mathrm{~mm}$, and period $p=33 \mathrm{~mm}$ for the meta-atom, \textcolor{black}{and the incident angle $\theta_{i}=73^{\circ}$}. For the simulations of anomalous deflection in Fig. 4 (D to F), the metasurface parameters are set as follows: to achieve a bending angle of $19^{\mathrm{o}}$, we set $a=31 \mathrm{~mm}, b=12.5 \mathrm{~mm}, t=$ $1 \mathrm{~mm}, m=5$, and $p=33 \mathrm{~mm}$; to achieve a bending angle of $34^{\mathrm{o}}$, we set $a=31 \mathrm{~mm}, b=12.5 \mathrm{~mm}, t=1 \mathrm{~mm}, m=3$, and $p=33 \mathrm{~mm}$; to achieve a bending angle of $59^{\mathrm{o}}$, we set $a=21 \mathrm{~mm}, b=12.5 \mathrm{~mm}, t=1 \mathrm{~mm}, m=3$, and $p=22 \mathrm{~mm}$. For the simulation of focusing in Fig. 4G, the metalens comprises 9 meta-atoms with dimensions $a=31 \mathrm{~mm}, b=12.5 \mathrm{~mm}, t=$ $1 \mathrm{~mm}$, and equal separation $p=33 \mathrm{~mm}$. For the simulations in Fig. 5, the Bessel beam has an aperture angle $\theta_{0}=35^{\circ}$. The meta-atom plates have $t=\lambda / 600$, $a=\lambda / 12$, and $b=\lambda / 3$ (length along $z$ direction). \textcolor{black}{All the meta-atoms in the simulations support a \textcolor{black}{dominant} dipole response at the working wavelegnth (see fig. S3)}.\\

\noindent\textbf{Experiment}\\
The experiments are conducted in a custom low-reflection environment (0.6 m × 0.52 m × 0.6 m) coated by sound
absorbing foams. The substrate and metasurfaces are fabricated using a three-dimensional printing technique (stereolithography) with photosensitive resin. For the phase gradient metasurfaces in Fig. 4 (D to F), we print 2 supercells, 3 supercells, and 5 supercells in a single-step modeling for achieving bending angle $19^{\mathrm{o}}$, $34^{\mathrm{o}}$, $59^{\mathrm{o}}$, respectively. The paramaters of the metasurface samples are $a=31 \mathrm{~mm}, b=12.5 \mathrm{~mm}, t=$ $1 \mathrm{~mm}, m=5$, and $p=33 \mathrm{~mm}$ for bending angle $19^{\mathrm{o}}$; $a=31 \mathrm{~mm}, b=12.5 \mathrm{~mm}, t=1 \mathrm{~mm}, m=3$, and $p=33 \mathrm{~mm}$ for bending angle $34^{\mathrm{o}}$; $a=21 \mathrm{~mm}, b=12.5 \mathrm{~mm}, t=1 \mathrm{~mm}, m=3$, and $p=22 \mathrm{~mm}$ for bending angle $59^{\mathrm{o}}$. The metalens sample comprises 9 meta-atoms with dimensions $a=31 \mathrm{~mm}, b=12.5 \mathrm{~mm}, t=$ $1 \mathrm{~mm}$, and equal separation $p=33 \mathrm{~mm}$. A sound and vibration module controlled by the host computer is used for signal generation and data acquisition. The sound field above the substrate is measured by a 1/4-inch microphone with a built-in preamplifie.\\

\noindent\textbf{{Acknowledgments}}\\
\noindent The work described in this paper was supported by grants from National Natural Science Foundation of China (No. 12322416) and Research Grants Council of the Hong Kong Special Administrative Region, China (Project No. AoE/P-502/20).\\

\noindent\textbf{Author contributions}\\
\noindent S.W. conceived the idea. W.X. performed the numerical simulations and analytical calculations. W.X., W.K., and S.H. conducted the experiment. W.X., W.K., S.H., S.L., D.P.T., and S.W. analyzed the results. W.X. and S.W. wrote the manuscript with input from all authors. S.L. and S.W. supervised the project.\\


\noindent\textbf{\large{References}}\\
\vspace{-1.05in}
\bibliography{apssamp}

\begin{thebibliography}{51}%
\makeatletter
\providecommand \@ifxundefined [1]{%
 \@ifx{#1\undefined}
}%
\providecommand \@ifnum [1]{%
 \ifnum #1\expandafter \@firstoftwo
 \else \expandafter \@secondoftwo
 \fi
}%
\providecommand \@ifx [1]{%
 \ifx #1\expandafter \@firstoftwo
 \else \expandafter \@secondoftwo
 \fi
}%
\providecommand \natexlab [1]{#1}%
\providecommand \enquote  [1]{``#1''}%
\providecommand \bibnamefont  [1]{#1}%
\providecommand \bibfnamefont [1]{#1}%
\providecommand \citenamefont [1]{#1}%
\providecommand \href@noop [0]{\@secondoftwo}%
\providecommand \href [0]{\begingroup \@sanitize@url \@href}%
\providecommand \@href[1]{\@@startlink{#1}\@@href}%
\providecommand \@@href[1]{\endgroup#1\@@endlink}%
\providecommand \@sanitize@url [0]{\catcode `\\12\catcode `\$12\catcode `\&12\catcode `\#12\catcode `\^12\catcode `\_12\catcode `\%12\relax}%
\providecommand \@@startlink[1]{}%
\providecommand \@@endlink[0]{}%
\providecommand \url  [0]{\begingroup\@sanitize@url \@url }%
\providecommand \@url [1]{\endgroup\@href {#1}{\urlprefix }}%
\providecommand \urlprefix  [0]{URL }%
\providecommand \Eprint [0]{\href }%
\providecommand \doibase [0]{https://doi.org/}%
\providecommand \selectlanguage [0]{\@gobble}%
\providecommand \bibinfo  [0]{\@secondoftwo}%
\providecommand \bibfield  [0]{\@secondoftwo}%
\providecommand \translation [1]{[#1]}%
\providecommand \BibitemOpen [0]{}%
\providecommand \bibitemStop [0]{}%
\providecommand \bibitemNoStop [0]{.\EOS\space}%
\providecommand \EOS [0]{\spacefactor3000\relax}%
\providecommand \BibitemShut  [1]{\csname bibitem#1\endcsname}%
\let\auto@bib@innerbib\@empty
\bibitem [{\citenamefont {Berry}(1984)}]{1berry1984quantal}%
  \BibitemOpen
  \bibfield  {author} {\bibinfo {author} {\bibfnamefont {M.~V.}\ \bibnamefont {Berry}},\ }\href@noop {} {\bibfield  {journal} {\bibinfo  {journal} {Proc. R. Soc. A: Math. Phys. Eng. Sci.}\ }\textbf {\bibinfo {volume} {392}},\ \bibinfo {pages} {45} (\bibinfo {year} {1984})}\BibitemShut {NoStop}%
\bibitem [{\citenamefont {Cisowski}\ \emph {et~al.}(2022)\citenamefont {Cisowski}, \citenamefont {G{\"o}tte},\ and\ \citenamefont {Franke-Arnold}}]{2gonoskov2022charged}%
  \BibitemOpen
  \bibfield  {author} {\bibinfo {author} {\bibfnamefont {C.}~\bibnamefont {Cisowski}}, \bibinfo {author} {\bibfnamefont {J.}~\bibnamefont {G{\"o}tte}},\ and\ \bibinfo {author} {\bibfnamefont {S.}~\bibnamefont {Franke-Arnold}},\ }\href@noop {} {\bibfield  {journal} {\bibinfo  {journal} {Rev. Mod. Phys.}\ }\textbf {\bibinfo {volume} {94}},\ \bibinfo {pages} {031001} (\bibinfo {year} {2022})}\BibitemShut {NoStop}%
\bibitem [{\citenamefont {Cohen}\ \emph {et~al.}(2019)\citenamefont {Cohen}, \citenamefont {Larocque}, \citenamefont {Bouchard}, \citenamefont {Nejadsattari}, \citenamefont {Gefen},\ and\ \citenamefont {Karimi}}]{58cohen2019geometric}%
  \BibitemOpen
  \bibfield  {author} {\bibinfo {author} {\bibfnamefont {E.}~\bibnamefont {Cohen}}, \bibinfo {author} {\bibfnamefont {H.}~\bibnamefont {Larocque}}, \bibinfo {author} {\bibfnamefont {F.}~\bibnamefont {Bouchard}}, \bibinfo {author} {\bibfnamefont {F.}~\bibnamefont {Nejadsattari}}, \bibinfo {author} {\bibfnamefont {Y.}~\bibnamefont {Gefen}},\ and\ \bibinfo {author} {\bibfnamefont {E.}~\bibnamefont {Karimi}},\ }\href@noop {} {\bibfield  {journal} {\bibinfo  {journal} {Nat. Rev. Phys.}\ }\textbf {\bibinfo {volume} {1}},\ \bibinfo {pages} {437} (\bibinfo {year} {2019})}\BibitemShut {NoStop}%
\bibitem [{\citenamefont {Simon}(1983)}]{3PhysRevLett.51.2167}%
  \BibitemOpen
  \bibfield  {author} {\bibinfo {author} {\bibfnamefont {B.}~\bibnamefont {Simon}},\ }\href@noop {} {\bibfield  {journal} {\bibinfo  {journal} {Phys. Rev. Lett.}\ }\textbf {\bibinfo {volume} {51}},\ \bibinfo {pages} {2167} (\bibinfo {year} {1983})}\BibitemShut {NoStop}%
\bibitem [{\citenamefont {Aharonov}\ and\ \citenamefont {Bohm}(1959)}]{4PhysRev.115.485}%
  \BibitemOpen
  \bibfield  {author} {\bibinfo {author} {\bibfnamefont {Y.}~\bibnamefont {Aharonov}}\ and\ \bibinfo {author} {\bibfnamefont {D.}~\bibnamefont {Bohm}},\ }\href@noop {} {\bibfield  {journal} {\bibinfo  {journal} {Phys. Rev.}\ }\textbf {\bibinfo {volume} {115}},\ \bibinfo {pages} {485} (\bibinfo {year} {1959})}\BibitemShut {NoStop}%
\bibitem [{\citenamefont {Thouless}\ \emph {et~al.}(1982)\citenamefont {Thouless}, \citenamefont {Kohmoto}, \citenamefont {Nightingale},\ and\ \citenamefont {den Nijs}}]{5PhysRevLett.49.405}%
  \BibitemOpen
  \bibfield  {author} {\bibinfo {author} {\bibfnamefont {D.~J.}\ \bibnamefont {Thouless}}, \bibinfo {author} {\bibfnamefont {M.}~\bibnamefont {Kohmoto}}, \bibinfo {author} {\bibfnamefont {M.~P.}\ \bibnamefont {Nightingale}},\ and\ \bibinfo {author} {\bibfnamefont {M.}~\bibnamefont {den Nijs}},\ }\href@noop {} {\bibfield  {journal} {\bibinfo  {journal} {Phys. Rev. Lett.}\ }\textbf {\bibinfo {volume} {49}},\ \bibinfo {pages} {405} (\bibinfo {year} {1982})}\BibitemShut {NoStop}%
\bibitem [{\citenamefont {Kane}\ and\ \citenamefont {Mele}(2005)}]{6PhysRevLett.95.226801}%
  \BibitemOpen
  \bibfield  {author} {\bibinfo {author} {\bibfnamefont {C.~L.}\ \bibnamefont {Kane}}\ and\ \bibinfo {author} {\bibfnamefont {E.~J.}\ \bibnamefont {Mele}},\ }\href@noop {} {\bibfield  {journal} {\bibinfo  {journal} {Phys. Rev. Lett.}\ }\textbf {\bibinfo {volume} {95}},\ \bibinfo {pages} {226801} (\bibinfo {year} {2005})}\BibitemShut {NoStop}%
\bibitem [{\citenamefont {Ozawa}\ \emph {et~al.}(2019)\citenamefont {Ozawa}, \citenamefont {Price}, \citenamefont {Amo}, \citenamefont {Goldman}, \citenamefont {Hafezi}, \citenamefont {Lu}, \citenamefont {Rechtsman}, \citenamefont {Schuster}, \citenamefont {Simon}, \citenamefont {Zilberberg},\ and\ \citenamefont {Carusotto}}]{9RevModPhys.91.015006}%
  \BibitemOpen
  \bibfield  {author} {\bibinfo {author} {\bibfnamefont {T.}~\bibnamefont {Ozawa}}, \bibinfo {author} {\bibfnamefont {H.~M.}\ \bibnamefont {Price}}, \bibinfo {author} {\bibfnamefont {A.}~\bibnamefont {Amo}}, \bibinfo {author} {\bibfnamefont {N.}~\bibnamefont {Goldman}}, \bibinfo {author} {\bibfnamefont {M.}~\bibnamefont {Hafezi}}, \bibinfo {author} {\bibfnamefont {L.}~\bibnamefont {Lu}}, \bibinfo {author} {\bibfnamefont {M.~C.}\ \bibnamefont {Rechtsman}}, \bibinfo {author} {\bibfnamefont {D.}~\bibnamefont {Schuster}}, \bibinfo {author} {\bibfnamefont {J.}~\bibnamefont {Simon}}, \bibinfo {author} {\bibfnamefont {O.}~\bibnamefont {Zilberberg}},\ and\ \bibinfo {author} {\bibfnamefont {I.}~\bibnamefont {Carusotto}},\ }\href@noop {} {\bibfield  {journal} {\bibinfo  {journal} {Rev. Mod. Phys.}\ }\textbf {\bibinfo {volume} {91}},\ \bibinfo {pages} {015006} (\bibinfo {year} {2019})}\BibitemShut {NoStop}%
\bibitem [{\citenamefont {Xue}\ \emph {et~al.}(2022)\citenamefont {Xue}, \citenamefont {Yang},\ and\ \citenamefont {Zhang}}]{10xue2022topological}%
  \BibitemOpen
  \bibfield  {author} {\bibinfo {author} {\bibfnamefont {H.}~\bibnamefont {Xue}}, \bibinfo {author} {\bibfnamefont {Y.}~\bibnamefont {Yang}},\ and\ \bibinfo {author} {\bibfnamefont {B.}~\bibnamefont {Zhang}},\ }\href@noop {} {\bibfield  {journal} {\bibinfo  {journal} {Nat. Rev. Mater.}\ }\textbf {\bibinfo {volume} {7}},\ \bibinfo {pages} {974} (\bibinfo {year} {2022})}\BibitemShut {NoStop}%
\bibitem [{\citenamefont {Wang}\ \emph {et~al.}(2009)\citenamefont {Wang}, \citenamefont {Chong}, \citenamefont {Joannopoulos},\ and\ \citenamefont {Solja{\v{c}}i{\'c}}}]{11wang2009observation}%
  \BibitemOpen
  \bibfield  {author} {\bibinfo {author} {\bibfnamefont {Z.}~\bibnamefont {Wang}}, \bibinfo {author} {\bibfnamefont {Y.}~\bibnamefont {Chong}}, \bibinfo {author} {\bibfnamefont {J.~D.}\ \bibnamefont {Joannopoulos}},\ and\ \bibinfo {author} {\bibfnamefont {M.}~\bibnamefont {Solja{\v{c}}i{\'c}}},\ }\href@noop {} {\bibfield  {journal} {\bibinfo  {journal} {Nature}\ }\textbf {\bibinfo {volume} {461}},\ \bibinfo {pages} {772} (\bibinfo {year} {2009})}\BibitemShut {NoStop}%
\bibitem [{\citenamefont {Yang}\ \emph {et~al.}(2020)\citenamefont {Yang}, \citenamefont {Yamagami}, \citenamefont {Yu}, \citenamefont {Pitchappa}, \citenamefont {Webber}, \citenamefont {Zhang}, \citenamefont {Fujita}, \citenamefont {Nagatsuma},\ and\ \citenamefont {Singh}}]{12yang2020terahertz}%
  \BibitemOpen
  \bibfield  {author} {\bibinfo {author} {\bibfnamefont {Y.}~\bibnamefont {Yang}}, \bibinfo {author} {\bibfnamefont {Y.}~\bibnamefont {Yamagami}}, \bibinfo {author} {\bibfnamefont {X.}~\bibnamefont {Yu}}, \bibinfo {author} {\bibfnamefont {P.}~\bibnamefont {Pitchappa}}, \bibinfo {author} {\bibfnamefont {J.}~\bibnamefont {Webber}}, \bibinfo {author} {\bibfnamefont {B.}~\bibnamefont {Zhang}}, \bibinfo {author} {\bibfnamefont {M.}~\bibnamefont {Fujita}}, \bibinfo {author} {\bibfnamefont {T.}~\bibnamefont {Nagatsuma}},\ and\ \bibinfo {author} {\bibfnamefont {R.}~\bibnamefont {Singh}},\ }\href@noop {} {\bibfield  {journal} {\bibinfo  {journal} {Nat. Photonics}\ }\textbf {\bibinfo {volume} {14}},\ \bibinfo {pages} {446} (\bibinfo {year} {2020})}\BibitemShut {NoStop}%
\bibitem [{\citenamefont {Bandres}\ \emph {et~al.}(2018)\citenamefont {Bandres}, \citenamefont {Wittek}, \citenamefont {Harari}, \citenamefont {Parto}, \citenamefont {Ren}, \citenamefont {Segev}, \citenamefont {Christodoulides},\ and\ \citenamefont {Khajavikhan}}]{13bandres2018topological}%
  \BibitemOpen
  \bibfield  {author} {\bibinfo {author} {\bibfnamefont {M.~A.}\ \bibnamefont {Bandres}}, \bibinfo {author} {\bibfnamefont {S.}~\bibnamefont {Wittek}}, \bibinfo {author} {\bibfnamefont {G.}~\bibnamefont {Harari}}, \bibinfo {author} {\bibfnamefont {M.}~\bibnamefont {Parto}}, \bibinfo {author} {\bibfnamefont {J.}~\bibnamefont {Ren}}, \bibinfo {author} {\bibfnamefont {M.}~\bibnamefont {Segev}}, \bibinfo {author} {\bibfnamefont {D.~N.}\ \bibnamefont {Christodoulides}},\ and\ \bibinfo {author} {\bibfnamefont {M.}~\bibnamefont {Khajavikhan}},\ }\href@noop {} {\bibfield  {journal} {\bibinfo  {journal} {Science}\ }\textbf {\bibinfo {volume} {359}},\ \bibinfo {pages} {eaar4005} (\bibinfo {year} {2018})}\BibitemShut {NoStop}%
\bibitem [{\citenamefont {Yang}\ \emph {et~al.}(2022)\citenamefont {Yang}, \citenamefont {Li}, \citenamefont {Gao},\ and\ \citenamefont {Lu}}]{14yang2022topological}%
  \BibitemOpen
  \bibfield  {author} {\bibinfo {author} {\bibfnamefont {L.}~\bibnamefont {Yang}}, \bibinfo {author} {\bibfnamefont {G.}~\bibnamefont {Li}}, \bibinfo {author} {\bibfnamefont {X.}~\bibnamefont {Gao}},\ and\ \bibinfo {author} {\bibfnamefont {L.}~\bibnamefont {Lu}},\ }\href@noop {} {\bibfield  {journal} {\bibinfo  {journal} {Nat. Photonics}\ }\textbf {\bibinfo {volume} {16}},\ \bibinfo {pages} {279} (\bibinfo {year} {2022})}\BibitemShut {NoStop}%
\bibitem [{\citenamefont {Mittal}\ \emph {et~al.}(2018)\citenamefont {Mittal}, \citenamefont {Goldschmidt},\ and\ \citenamefont {Hafezi}}]{15mittal2018topological}%
  \BibitemOpen
  \bibfield  {author} {\bibinfo {author} {\bibfnamefont {S.}~\bibnamefont {Mittal}}, \bibinfo {author} {\bibfnamefont {E.~A.}\ \bibnamefont {Goldschmidt}},\ and\ \bibinfo {author} {\bibfnamefont {M.}~\bibnamefont {Hafezi}},\ }\href@noop {} {\bibfield  {journal} {\bibinfo  {journal} {Nature}\ }\textbf {\bibinfo {volume} {561}},\ \bibinfo {pages} {502} (\bibinfo {year} {2018})}\BibitemShut {NoStop}%
\bibitem [{\citenamefont {Barik}\ \emph {et~al.}(2018)\citenamefont {Barik}, \citenamefont {Karasahin}, \citenamefont {Flower}, \citenamefont {Cai}, \citenamefont {Miyake}, \citenamefont {DeGottardi}, \citenamefont {Hafezi},\ and\ \citenamefont {Waks}}]{16barik2018topological}%
  \BibitemOpen
  \bibfield  {author} {\bibinfo {author} {\bibfnamefont {S.}~\bibnamefont {Barik}}, \bibinfo {author} {\bibfnamefont {A.}~\bibnamefont {Karasahin}}, \bibinfo {author} {\bibfnamefont {C.}~\bibnamefont {Flower}}, \bibinfo {author} {\bibfnamefont {T.}~\bibnamefont {Cai}}, \bibinfo {author} {\bibfnamefont {H.}~\bibnamefont {Miyake}}, \bibinfo {author} {\bibfnamefont {W.}~\bibnamefont {DeGottardi}}, \bibinfo {author} {\bibfnamefont {M.}~\bibnamefont {Hafezi}},\ and\ \bibinfo {author} {\bibfnamefont {E.}~\bibnamefont {Waks}},\ }\href@noop {} {\bibfield  {journal} {\bibinfo  {journal} {Science}\ }\textbf {\bibinfo {volume} {359}},\ \bibinfo {pages} {666} (\bibinfo {year} {2018})}\BibitemShut {NoStop}%
\bibitem [{\citenamefont {Pancharatnam}(1956)}]{21pancharatnam1956generalized}%
  \BibitemOpen
  \bibfield  {author} {\bibinfo {author} {\bibfnamefont {S.}~\bibnamefont {Pancharatnam}},\ }\href@noop {} {\bibfield  {journal} {\bibinfo  {journal} {Proc. Indian Acad. Sci.}\ }\textbf {\bibinfo {volume} {44}},\ \bibinfo {pages} {247} (\bibinfo {year} {1956})}\BibitemShut {NoStop}%
\bibitem [{\citenamefont {Berry}(1987)}]{22berry1987adiabatic}%
  \BibitemOpen
  \bibfield  {author} {\bibinfo {author} {\bibfnamefont {M.~V.}\ \bibnamefont {Berry}},\ }\href@noop {} {\bibfield  {journal} {\bibinfo  {journal} {J. Mod. Opt.}\ }\textbf {\bibinfo {volume} {34}},\ \bibinfo {pages} {1401} (\bibinfo {year} {1987})}\BibitemShut {NoStop}%
\bibitem [{\citenamefont {Bomzon}\ \emph {et~al.}(2002)\citenamefont {Bomzon}, \citenamefont {Biener}, \citenamefont {Kleiner},\ and\ \citenamefont {Hasman}}]{31bomzon2002space}%
  \BibitemOpen
  \bibfield  {author} {\bibinfo {author} {\bibfnamefont {Z.}~\bibnamefont {Bomzon}}, \bibinfo {author} {\bibfnamefont {G.}~\bibnamefont {Biener}}, \bibinfo {author} {\bibfnamefont {V.}~\bibnamefont {Kleiner}},\ and\ \bibinfo {author} {\bibfnamefont {E.}~\bibnamefont {Hasman}},\ }\href@noop {} {\bibfield  {journal} {\bibinfo  {journal} {Opt. Lett.}\ }\textbf {\bibinfo {volume} {27}},\ \bibinfo {pages} {1141} (\bibinfo {year} {2002})}\BibitemShut {NoStop}%
\bibitem [{\citenamefont {Marrucci}\ \emph {et~al.}(2006)\citenamefont {Marrucci}, \citenamefont {Manzo},\ and\ \citenamefont {Paparo}}]{32marrucci2006optical}%
  \BibitemOpen
  \bibfield  {author} {\bibinfo {author} {\bibfnamefont {L.}~\bibnamefont {Marrucci}}, \bibinfo {author} {\bibfnamefont {C.}~\bibnamefont {Manzo}},\ and\ \bibinfo {author} {\bibfnamefont {D.}~\bibnamefont {Paparo}},\ }\href@noop {} {\bibfield  {journal} {\bibinfo  {journal} {Phys. Rev. Lett.}\ }\textbf {\bibinfo {volume} {96}},\ \bibinfo {pages} {163905} (\bibinfo {year} {2006})}\BibitemShut {NoStop}%
\bibitem [{\citenamefont {Xiao}\ and\ \citenamefont {Wang}(2024)}]{56Xiao:24}%
  \BibitemOpen
  \bibfield  {author} {\bibinfo {author} {\bibfnamefont {W.}~\bibnamefont {Xiao}}\ and\ \bibinfo {author} {\bibfnamefont {S.}~\bibnamefont {Wang}},\ }\href@noop {} {\bibfield  {journal} {\bibinfo  {journal} {Opt. Lett.}\ }\textbf {\bibinfo {volume} {49}},\ \bibinfo {pages} {1915} (\bibinfo {year} {2024})}\BibitemShut {NoStop}%
\bibitem [{\citenamefont {Wang}\ \emph {et~al.}(2018)\citenamefont {Wang}, \citenamefont {Wu}, \citenamefont {Su}, \citenamefont {Lai}, \citenamefont {Chen}, \citenamefont {Kuo}, \citenamefont {Chen}, \citenamefont {Chen}, \citenamefont {Huang}, \citenamefont {Wang}, \citenamefont {Lin}, \citenamefont {Kuan}, \citenamefont {Li}, \citenamefont {Wang}, \citenamefont {Zhu},\ and\ \citenamefont {Tsai}}]{33wang2018broadband}%
  \BibitemOpen
  \bibfield  {author} {\bibinfo {author} {\bibfnamefont {S.}~\bibnamefont {Wang}}, \bibinfo {author} {\bibfnamefont {P.~C.}\ \bibnamefont {Wu}}, \bibinfo {author} {\bibfnamefont {V.-C.}\ \bibnamefont {Su}}, \bibinfo {author} {\bibfnamefont {Y.-C.}\ \bibnamefont {Lai}}, \bibinfo {author} {\bibfnamefont {M.-K.}\ \bibnamefont {Chen}}, \bibinfo {author} {\bibfnamefont {H.~Y.}\ \bibnamefont {Kuo}}, \bibinfo {author} {\bibfnamefont {B.~H.}\ \bibnamefont {Chen}}, \bibinfo {author} {\bibfnamefont {Y.~H.}\ \bibnamefont {Chen}}, \bibinfo {author} {\bibfnamefont {T.-T.}\ \bibnamefont {Huang}}, \bibinfo {author} {\bibfnamefont {J.-H.}\ \bibnamefont {Wang}}, \bibinfo {author} {\bibfnamefont {R.-M.}\ \bibnamefont {Lin}}, \bibinfo {author} {\bibfnamefont {C.-H.}\ \bibnamefont {Kuan}}, \bibinfo {author} {\bibfnamefont {T.}~\bibnamefont {Li}}, \bibinfo {author} {\bibfnamefont {Z.}~\bibnamefont {Wang}}, \bibinfo {author} {\bibfnamefont {S.}~\bibnamefont {Zhu}},\ and\ \bibinfo {author} {\bibfnamefont {D.~P.}\ \bibnamefont
  {Tsai}},\ }\href@noop {} {\bibfield  {journal} {\bibinfo  {journal} {Nat. Nanotechnol.}\ }\textbf {\bibinfo {volume} {13}},\ \bibinfo {pages} {227} (\bibinfo {year} {2018})}\BibitemShut {NoStop}%
\bibitem [{\citenamefont {Chen}\ \emph {et~al.}(2018)\citenamefont {Chen}, \citenamefont {Zhu}, \citenamefont {Sanjeev}, \citenamefont {Khorasaninejad}, \citenamefont {Shi}, \citenamefont {Lee},\ and\ \citenamefont {Capasso}}]{34chen2018broadband}%
  \BibitemOpen
  \bibfield  {author} {\bibinfo {author} {\bibfnamefont {W.~T.}\ \bibnamefont {Chen}}, \bibinfo {author} {\bibfnamefont {A.~Y.}\ \bibnamefont {Zhu}}, \bibinfo {author} {\bibfnamefont {V.}~\bibnamefont {Sanjeev}}, \bibinfo {author} {\bibfnamefont {M.}~\bibnamefont {Khorasaninejad}}, \bibinfo {author} {\bibfnamefont {Z.}~\bibnamefont {Shi}}, \bibinfo {author} {\bibfnamefont {E.}~\bibnamefont {Lee}},\ and\ \bibinfo {author} {\bibfnamefont {F.}~\bibnamefont {Capasso}},\ }\href@noop {} {\bibfield  {journal} {\bibinfo  {journal} {Nat. Nanotechnol.}\ }\textbf {\bibinfo {volume} {13}},\ \bibinfo {pages} {220} (\bibinfo {year} {2018})}\BibitemShut {NoStop}%
\bibitem [{\citenamefont {Zheng}\ \emph {et~al.}(2015)\citenamefont {Zheng}, \citenamefont {M{\"u}hlenbernd}, \citenamefont {Kenney}, \citenamefont {Li}, \citenamefont {Zentgraf},\ and\ \citenamefont {Zhang}}]{35zheng2015metasurface}%
  \BibitemOpen
  \bibfield  {author} {\bibinfo {author} {\bibfnamefont {G.}~\bibnamefont {Zheng}}, \bibinfo {author} {\bibfnamefont {H.}~\bibnamefont {M{\"u}hlenbernd}}, \bibinfo {author} {\bibfnamefont {M.}~\bibnamefont {Kenney}}, \bibinfo {author} {\bibfnamefont {G.}~\bibnamefont {Li}}, \bibinfo {author} {\bibfnamefont {T.}~\bibnamefont {Zentgraf}},\ and\ \bibinfo {author} {\bibfnamefont {S.}~\bibnamefont {Zhang}},\ }\href@noop {} {\bibfield  {journal} {\bibinfo  {journal} {Nat. Nanotechnol.}\ }\textbf {\bibinfo {volume} {10}},\ \bibinfo {pages} {308} (\bibinfo {year} {2015})}\BibitemShut {NoStop}%
\bibitem [{\citenamefont {Li}\ \emph {et~al.}(2018)\citenamefont {Li}, \citenamefont {Kamin}, \citenamefont {Zheng}, \citenamefont {Neubrech}, \citenamefont {Zhang},\ and\ \citenamefont {Liu}}]{li2018addressable}%
  \BibitemOpen
  \bibfield  {author} {\bibinfo {author} {\bibfnamefont {J.}~\bibnamefont {Li}}, \bibinfo {author} {\bibfnamefont {S.}~\bibnamefont {Kamin}}, \bibinfo {author} {\bibfnamefont {G.}~\bibnamefont {Zheng}}, \bibinfo {author} {\bibfnamefont {F.}~\bibnamefont {Neubrech}}, \bibinfo {author} {\bibfnamefont {S.}~\bibnamefont {Zhang}},\ and\ \bibinfo {author} {\bibfnamefont {N.}~\bibnamefont {Liu}},\ }\href@noop {} {\bibfield  {journal} {\bibinfo  {journal} {Sci. Adv.}\ }\textbf {\bibinfo {volume} {4}},\ \bibinfo {pages} {eaar6768} (\bibinfo {year} {2018})}\BibitemShut {NoStop}%
\bibitem [{\citenamefont {Li}\ \emph {et~al.}(2015)\citenamefont {Li}, \citenamefont {Chen}, \citenamefont {Pholchai}, \citenamefont {Reineke}, \citenamefont {Wong}, \citenamefont {Pun}, \citenamefont {Cheah}, \citenamefont {Zentgraf},\ and\ \citenamefont {Zhang}}]{36li2015continuous}%
  \BibitemOpen
  \bibfield  {author} {\bibinfo {author} {\bibfnamefont {G.}~\bibnamefont {Li}}, \bibinfo {author} {\bibfnamefont {S.}~\bibnamefont {Chen}}, \bibinfo {author} {\bibfnamefont {N.}~\bibnamefont {Pholchai}}, \bibinfo {author} {\bibfnamefont {B.}~\bibnamefont {Reineke}}, \bibinfo {author} {\bibfnamefont {P.~W.~H.}\ \bibnamefont {Wong}}, \bibinfo {author} {\bibfnamefont {E.~Y.~B.}\ \bibnamefont {Pun}}, \bibinfo {author} {\bibfnamefont {K.~W.}\ \bibnamefont {Cheah}}, \bibinfo {author} {\bibfnamefont {T.}~\bibnamefont {Zentgraf}},\ and\ \bibinfo {author} {\bibfnamefont {S.}~\bibnamefont {Zhang}},\ }\href@noop {} {\bibfield  {journal} {\bibinfo  {journal} {Nat. Mater.}\ }\textbf {\bibinfo {volume} {14}},\ \bibinfo {pages} {607} (\bibinfo {year} {2015})}\BibitemShut {NoStop}%
\bibitem [{\citenamefont {Liu}\ \emph {et~al.}(2021)\citenamefont {Liu}, \citenamefont {Su}, \citenamefont {Zeng}, \citenamefont {Wang}, \citenamefont {Huang},\ and\ \citenamefont {Zhang}}]{43liu2021acoustic}%
  \BibitemOpen
  \bibfield  {author} {\bibinfo {author} {\bibfnamefont {B.}~\bibnamefont {Liu}}, \bibinfo {author} {\bibfnamefont {Z.}~\bibnamefont {Su}}, \bibinfo {author} {\bibfnamefont {Y.}~\bibnamefont {Zeng}}, \bibinfo {author} {\bibfnamefont {Y.}~\bibnamefont {Wang}}, \bibinfo {author} {\bibfnamefont {L.}~\bibnamefont {Huang}},\ and\ \bibinfo {author} {\bibfnamefont {S.}~\bibnamefont {Zhang}},\ }\href@noop {} {\bibfield  {journal} {\bibinfo  {journal} {New J. Phys.}\ }\textbf {\bibinfo {volume} {23}},\ \bibinfo {pages} {113026} (\bibinfo {year} {2021})}\BibitemShut {NoStop}%
\bibitem [{\citenamefont {Liu}\ \emph {et~al.}(2022)\citenamefont {Liu}, \citenamefont {Zhou}, \citenamefont {Wang}, \citenamefont {Zentgraf}, \citenamefont {Li},\ and\ \citenamefont {Huang}}]{44liu2022experimental}%
  \BibitemOpen
  \bibfield  {author} {\bibinfo {author} {\bibfnamefont {B.}~\bibnamefont {Liu}}, \bibinfo {author} {\bibfnamefont {Z.}~\bibnamefont {Zhou}}, \bibinfo {author} {\bibfnamefont {Y.}~\bibnamefont {Wang}}, \bibinfo {author} {\bibfnamefont {T.}~\bibnamefont {Zentgraf}}, \bibinfo {author} {\bibfnamefont {Y.}~\bibnamefont {Li}},\ and\ \bibinfo {author} {\bibfnamefont {L.}~\bibnamefont {Huang}},\ }\href@noop {} {\bibfield  {journal} {\bibinfo  {journal} {Appl. Phys. Lett.}\ }\textbf {\bibinfo {volume} {120}} (\bibinfo {year} {2022})}\BibitemShut {NoStop}%
\bibitem [{\citenamefont {Zhang}\ \emph {et~al.}(2023)\citenamefont {Zhang}, \citenamefont {Li}, \citenamefont {Dong}, \citenamefont {Xue}, \citenamefont {You}, \citenamefont {Liu}, \citenamefont {Gao}, \citenamefont {Jiang}, \citenamefont {Chen}, \citenamefont {Xu},\ and\ \citenamefont {Fu}}]{45zhang2023geometric}%
  \BibitemOpen
  \bibfield  {author} {\bibinfo {author} {\bibfnamefont {K.}~\bibnamefont {Zhang}}, \bibinfo {author} {\bibfnamefont {X.}~\bibnamefont {Li}}, \bibinfo {author} {\bibfnamefont {D.}~\bibnamefont {Dong}}, \bibinfo {author} {\bibfnamefont {M.}~\bibnamefont {Xue}}, \bibinfo {author} {\bibfnamefont {W.-L.}\ \bibnamefont {You}}, \bibinfo {author} {\bibfnamefont {Y.}~\bibnamefont {Liu}}, \bibinfo {author} {\bibfnamefont {L.}~\bibnamefont {Gao}}, \bibinfo {author} {\bibfnamefont {J.-H.}\ \bibnamefont {Jiang}}, \bibinfo {author} {\bibfnamefont {H.}~\bibnamefont {Chen}}, \bibinfo {author} {\bibfnamefont {Y.}~\bibnamefont {Xu}},\ and\ \bibinfo {author} {\bibfnamefont {Y.}~\bibnamefont {Fu}},\ }\href@noop {} {\bibfield  {journal} {\bibinfo  {journal} {Adv. Sci}\ }\textbf {\bibinfo {volume} {10}},\ \bibinfo {pages} {2304992} (\bibinfo {year} {2023})}\BibitemShut {NoStop}%
\bibitem [{\citenamefont {Chen}\ \emph {et~al.}(2024)\citenamefont {Chen}, \citenamefont {Li}, \citenamefont {Li}, \citenamefont {Xue}, \citenamefont {Shi}, \citenamefont {Dong}, \citenamefont {Xu}, \citenamefont {Liu},\ and\ \citenamefont {Fu}}]{chen2024super}%
  \BibitemOpen
  \bibfield  {author} {\bibinfo {author} {\bibfnamefont {C.}~\bibnamefont {Chen}}, \bibinfo {author} {\bibfnamefont {X.}~\bibnamefont {Li}}, \bibinfo {author} {\bibfnamefont {W.}~\bibnamefont {Li}}, \bibinfo {author} {\bibfnamefont {M.}~\bibnamefont {Xue}}, \bibinfo {author} {\bibfnamefont {Y.}~\bibnamefont {Shi}}, \bibinfo {author} {\bibfnamefont {D.}~\bibnamefont {Dong}}, \bibinfo {author} {\bibfnamefont {Y.}~\bibnamefont {Xu}}, \bibinfo {author} {\bibfnamefont {Y.}~\bibnamefont {Liu}},\ and\ \bibinfo {author} {\bibfnamefont {Y.}~\bibnamefont {Fu}},\ }\href@noop {} {\bibfield  {journal} {\bibinfo  {journal} {Nat. Commun.}\ }\textbf {\bibinfo {volume} {15}},\ \bibinfo {pages} {8391} (\bibinfo {year} {2024})}\BibitemShut {NoStop}%
\bibitem [{\citenamefont {Shi}\ \emph {et~al.}(2019)\citenamefont {Shi}, \citenamefont {Zhao}, \citenamefont {Long}, \citenamefont {Yang}, \citenamefont {Wang}, \citenamefont {Chen}, \citenamefont {Ren},\ and\ \citenamefont {Zhang}}]{47shi2019observation}%
  \BibitemOpen
  \bibfield  {author} {\bibinfo {author} {\bibfnamefont {C.}~\bibnamefont {Shi}}, \bibinfo {author} {\bibfnamefont {R.}~\bibnamefont {Zhao}}, \bibinfo {author} {\bibfnamefont {Y.}~\bibnamefont {Long}}, \bibinfo {author} {\bibfnamefont {S.}~\bibnamefont {Yang}}, \bibinfo {author} {\bibfnamefont {Y.}~\bibnamefont {Wang}}, \bibinfo {author} {\bibfnamefont {H.}~\bibnamefont {Chen}}, \bibinfo {author} {\bibfnamefont {J.}~\bibnamefont {Ren}},\ and\ \bibinfo {author} {\bibfnamefont {X.}~\bibnamefont {Zhang}},\ }\href@noop {} {\bibfield  {journal} {\bibinfo  {journal} {Natl. Sci. Rev.}\ }\textbf {\bibinfo {volume} {6}},\ \bibinfo {pages} {707} (\bibinfo {year} {2019})}\BibitemShut {NoStop}%
\bibitem [{\citenamefont {Bliokh}\ and\ \citenamefont {Nori}(2019)}]{46PhysRevB.99.174310}%
  \BibitemOpen
  \bibfield  {author} {\bibinfo {author} {\bibfnamefont {K.~Y.}\ \bibnamefont {Bliokh}}\ and\ \bibinfo {author} {\bibfnamefont {F.}~\bibnamefont {Nori}},\ }\href@noop {} {\bibfield  {journal} {\bibinfo  {journal} {Phys. Rev. B}\ }\textbf {\bibinfo {volume} {99}},\ \bibinfo {pages} {174310} (\bibinfo {year} {2019})}\BibitemShut {NoStop}%
\bibitem [{\citenamefont {Long}\ \emph {et~al.}(2020)\citenamefont {Long}, \citenamefont {Ge}, \citenamefont {Zhang}, \citenamefont {Xu}, \citenamefont {Ren}, \citenamefont {Lu}, \citenamefont {Bao}, \citenamefont {Chen},\ and\ \citenamefont {Chen}}]{64long2020symmetry}%
  \BibitemOpen
  \bibfield  {author} {\bibinfo {author} {\bibfnamefont {Y.}~\bibnamefont {Long}}, \bibinfo {author} {\bibfnamefont {H.}~\bibnamefont {Ge}}, \bibinfo {author} {\bibfnamefont {D.}~\bibnamefont {Zhang}}, \bibinfo {author} {\bibfnamefont {X.}~\bibnamefont {Xu}}, \bibinfo {author} {\bibfnamefont {J.}~\bibnamefont {Ren}}, \bibinfo {author} {\bibfnamefont {M.-H.}\ \bibnamefont {Lu}}, \bibinfo {author} {\bibfnamefont {M.}~\bibnamefont {Bao}}, \bibinfo {author} {\bibfnamefont {H.}~\bibnamefont {Chen}},\ and\ \bibinfo {author} {\bibfnamefont {Y.-F.}\ \bibnamefont {Chen}},\ }\href@noop {} {\bibfield  {journal} {\bibinfo  {journal} {Natl. Sci. Rev.}\ }\textbf {\bibinfo {volume} {7}},\ \bibinfo {pages} {1024} (\bibinfo {year} {2020})}\BibitemShut {NoStop}%
\bibitem [{\citenamefont {Wang}\ \emph {et~al.}(2021)\citenamefont {Wang}, \citenamefont {Zhang}, \citenamefont {Wang}, \citenamefont {Tong}, \citenamefont {Li},\ and\ \citenamefont {Ma}}]{48wang2021spin}%
  \BibitemOpen
  \bibfield  {author} {\bibinfo {author} {\bibfnamefont {S.}~\bibnamefont {Wang}}, \bibinfo {author} {\bibfnamefont {G.}~\bibnamefont {Zhang}}, \bibinfo {author} {\bibfnamefont {X.}~\bibnamefont {Wang}}, \bibinfo {author} {\bibfnamefont {Q.}~\bibnamefont {Tong}}, \bibinfo {author} {\bibfnamefont {J.}~\bibnamefont {Li}},\ and\ \bibinfo {author} {\bibfnamefont {G.}~\bibnamefont {Ma}},\ }\href@noop {} {\bibfield  {journal} {\bibinfo  {journal} {Nat. Commun.}\ }\textbf {\bibinfo {volume} {12}},\ \bibinfo {pages} {6125} (\bibinfo {year} {2021})}\BibitemShut {NoStop}%
\bibitem [{\citenamefont {Muelas-Hurtado}\ \emph {et~al.}(2022)\citenamefont {Muelas-Hurtado}, \citenamefont {Volke-Sep\'ulveda}, \citenamefont {Ealo}, \citenamefont {Nori}, \citenamefont {Alonso}, \citenamefont {Bliokh},\ and\ \citenamefont {Brasselet}}]{69PhysRevLett.129.204301}%
  \BibitemOpen
  \bibfield  {author} {\bibinfo {author} {\bibfnamefont {R.~D.}\ \bibnamefont {Muelas-Hurtado}}, \bibinfo {author} {\bibfnamefont {K.}~\bibnamefont {Volke-Sep\'ulveda}}, \bibinfo {author} {\bibfnamefont {J.~L.}\ \bibnamefont {Ealo}}, \bibinfo {author} {\bibfnamefont {F.}~\bibnamefont {Nori}}, \bibinfo {author} {\bibfnamefont {M.~A.}\ \bibnamefont {Alonso}}, \bibinfo {author} {\bibfnamefont {K.~Y.}\ \bibnamefont {Bliokh}},\ and\ \bibinfo {author} {\bibfnamefont {E.}~\bibnamefont {Brasselet}},\ }\href@noop {} {\bibfield  {journal} {\bibinfo  {journal} {Phys. Rev. Lett.}\ }\textbf {\bibinfo {volume} {129}},\ \bibinfo {pages} {204301} (\bibinfo {year} {2022})}\BibitemShut {NoStop}%
\bibitem [{\citenamefont {Alha\"{\i}tz}\ \emph {et~al.}(2023)\citenamefont {Alha\"{\i}tz}, \citenamefont {Brunet}, \citenamefont {Arist\'egui}, \citenamefont {Poncelet},\ and\ \citenamefont {Baresch}}]{64PhysRevLett.131.114001}%
  \BibitemOpen
  \bibfield  {author} {\bibinfo {author} {\bibfnamefont {L.}~\bibnamefont {Alha\"{\i}tz}}, \bibinfo {author} {\bibfnamefont {T.}~\bibnamefont {Brunet}}, \bibinfo {author} {\bibfnamefont {C.}~\bibnamefont {Arist\'egui}}, \bibinfo {author} {\bibfnamefont {O.}~\bibnamefont {Poncelet}},\ and\ \bibinfo {author} {\bibfnamefont {D.}~\bibnamefont {Baresch}},\ }\href@noop {} {\bibfield  {journal} {\bibinfo  {journal} {Phys. Rev. Lett.}\ }\textbf {\bibinfo {volume} {131}},\ \bibinfo {pages} {114001} (\bibinfo {year} {2023})}\BibitemShut {NoStop}%
\bibitem [{\citenamefont {Tong}\ and\ \citenamefont {Wang}(2025)}]{70tong2023topological}%
  \BibitemOpen
  \bibfield  {author} {\bibinfo {author} {\bibfnamefont {Q.}~\bibnamefont {Tong}}\ and\ \bibinfo {author} {\bibfnamefont {S.}~\bibnamefont {Wang}},\ }\href@noop {} {\bibfield  {journal} {\bibinfo  {journal} {New J. Phys.}\ }\textbf {\bibinfo {volume} {27}},\ \bibinfo {pages} {013020} (\bibinfo {year} {2025})}\BibitemShut {NoStop}%
\bibitem [{\citenamefont {Zhu}\ \emph {et~al.}(2011)\citenamefont {Zhu}, \citenamefont {Christensen}, \citenamefont {Jung}, \citenamefont {Martin-Moreno}, \citenamefont {Yin}, \citenamefont {Fok}, \citenamefont {Zhang},\ and\ \citenamefont {Garcia-Vidal}}]{zhu2011holey}%
  \BibitemOpen
  \bibfield  {author} {\bibinfo {author} {\bibfnamefont {J.}~\bibnamefont {Zhu}}, \bibinfo {author} {\bibfnamefont {J.}~\bibnamefont {Christensen}}, \bibinfo {author} {\bibfnamefont {J.}~\bibnamefont {Jung}}, \bibinfo {author} {\bibfnamefont {L.}~\bibnamefont {Martin-Moreno}}, \bibinfo {author} {\bibfnamefont {X.}~\bibnamefont {Yin}}, \bibinfo {author} {\bibfnamefont {L.}~\bibnamefont {Fok}}, \bibinfo {author} {\bibfnamefont {X.}~\bibnamefont {Zhang}},\ and\ \bibinfo {author} {\bibfnamefont {F.}~\bibnamefont {Garcia-Vidal}},\ }\href@noop {} {\bibfield  {journal} {\bibinfo  {journal} {Nat. Phys.}\ }\textbf {\bibinfo {volume} {7}},\ \bibinfo {pages} {52} (\bibinfo {year} {2011})}\BibitemShut {NoStop}%
\bibitem [{\citenamefont {Liu}\ \emph {et~al.}(2019)\citenamefont {Liu}, \citenamefont {Chen}, \citenamefont {Liang}, \citenamefont {Gao},\ and\ \citenamefont {Zhu}}]{50PhysRevApplied.11.034061}%
  \BibitemOpen
  \bibfield  {author} {\bibinfo {author} {\bibfnamefont {T.}~\bibnamefont {Liu}}, \bibinfo {author} {\bibfnamefont {F.}~\bibnamefont {Chen}}, \bibinfo {author} {\bibfnamefont {S.}~\bibnamefont {Liang}}, \bibinfo {author} {\bibfnamefont {H.}~\bibnamefont {Gao}},\ and\ \bibinfo {author} {\bibfnamefont {J.}~\bibnamefont {Zhu}},\ }\href@noop {} {\bibfield  {journal} {\bibinfo  {journal} {Phys. Rev. Appl.}\ }\textbf {\bibinfo {volume} {11}},\ \bibinfo {pages} {034061} (\bibinfo {year} {2019})}\BibitemShut {NoStop}%
\bibitem [{\citenamefont {Bliokh}\ and\ \citenamefont {Nori}(2015)}]{52bliokh2015transverse}%
  \BibitemOpen
  \bibfield  {author} {\bibinfo {author} {\bibfnamefont {K.~Y.}\ \bibnamefont {Bliokh}}\ and\ \bibinfo {author} {\bibfnamefont {F.}~\bibnamefont {Nori}},\ }\href@noop {} {\bibfield  {journal} {\bibinfo  {journal} {Phys. Rep.}\ }\textbf {\bibinfo {volume} {592}},\ \bibinfo {pages} {1} (\bibinfo {year} {2015})}\BibitemShut {NoStop}%
\bibitem [{\citenamefont {Van~Mechelen}\ and\ \citenamefont {Jacob}(2016)}]{55van2016universal}%
  \BibitemOpen
  \bibfield  {author} {\bibinfo {author} {\bibfnamefont {T.}~\bibnamefont {Van~Mechelen}}\ and\ \bibinfo {author} {\bibfnamefont {Z.}~\bibnamefont {Jacob}},\ }\href@noop {} {\bibfield  {journal} {\bibinfo  {journal} {Optica}\ }\textbf {\bibinfo {volume} {3}},\ \bibinfo {pages} {118} (\bibinfo {year} {2016})}\BibitemShut {NoStop}%
\bibitem [{\citenamefont {Wang}\ \emph {et~al.}(2019)\citenamefont {Wang}, \citenamefont {Hou}, \citenamefont {Lu}, \citenamefont {Chen}, \citenamefont {Zhang},\ and\ \citenamefont {Chan}}]{swang19}%
  \BibitemOpen
  \bibfield  {author} {\bibinfo {author} {\bibfnamefont {S.}~\bibnamefont {Wang}}, \bibinfo {author} {\bibfnamefont {B.}~\bibnamefont {Hou}}, \bibinfo {author} {\bibfnamefont {W.}~\bibnamefont {Lu}}, \bibinfo {author} {\bibfnamefont {Y.}~\bibnamefont {Chen}}, \bibinfo {author} {\bibfnamefont {Z.~Q.}\ \bibnamefont {Zhang}},\ and\ \bibinfo {author} {\bibfnamefont {C.~T.}\ \bibnamefont {Chan}},\ }\href@noop {} {\bibfield  {journal} {\bibinfo  {journal} {Nat. Commun.}\ }\textbf {\bibinfo {volume} {10}},\ \bibinfo {pages} {832} (\bibinfo {year} {2019})}\BibitemShut {NoStop}%
\bibitem [{\citenamefont {Jackson}()}]{71jackson2012classical}%
  \BibitemOpen
  \bibfield  {author} {\bibinfo {author} {\bibfnamefont {J.~D.}\ \bibnamefont {Jackson}},\ }\href@noop {} {\emph {\bibinfo {title} {Classical Electrodynamics}}}\ (\bibinfo  {publisher} {John Wiley \& Sons, 2012})\BibitemShut {NoStop}%
\bibitem [{\citenamefont {Yu}\ \emph {et~al.}(2011)\citenamefont {Yu}, \citenamefont {Genevet}, \citenamefont {Kats}, \citenamefont {Aieta}, \citenamefont {Tetienne}, \citenamefont {Capasso},\ and\ \citenamefont {Gaburro}}]{60yuscience}%
  \BibitemOpen
  \bibfield  {author} {\bibinfo {author} {\bibfnamefont {N.}~\bibnamefont {Yu}}, \bibinfo {author} {\bibfnamefont {P.}~\bibnamefont {Genevet}}, \bibinfo {author} {\bibfnamefont {M.~A.}\ \bibnamefont {Kats}}, \bibinfo {author} {\bibfnamefont {F.}~\bibnamefont {Aieta}}, \bibinfo {author} {\bibfnamefont {J.-P.}\ \bibnamefont {Tetienne}}, \bibinfo {author} {\bibfnamefont {F.}~\bibnamefont {Capasso}},\ and\ \bibinfo {author} {\bibfnamefont {Z.}~\bibnamefont {Gaburro}},\ }\href@noop {} {\bibfield  {journal} {\bibinfo  {journal} {Science}\ }\textbf {\bibinfo {volume} {334}},\ \bibinfo {pages} {333} (\bibinfo {year} {2011})}\BibitemShut {NoStop}%
\bibitem [{\citenamefont {Assouar}\ \emph {et~al.}(2018)\citenamefont {Assouar}, \citenamefont {Liang}, \citenamefont {Wu}, \citenamefont {Li}, \citenamefont {Cheng},\ and\ \citenamefont {Jing}}]{assouar2018acoustic}%
  \BibitemOpen
  \bibfield  {author} {\bibinfo {author} {\bibfnamefont {B.}~\bibnamefont {Assouar}}, \bibinfo {author} {\bibfnamefont {B.}~\bibnamefont {Liang}}, \bibinfo {author} {\bibfnamefont {Y.}~\bibnamefont {Wu}}, \bibinfo {author} {\bibfnamefont {Y.}~\bibnamefont {Li}}, \bibinfo {author} {\bibfnamefont {J.-C.}\ \bibnamefont {Cheng}},\ and\ \bibinfo {author} {\bibfnamefont {Y.}~\bibnamefont {Jing}},\ }\href@noop {} {\bibfield  {journal} {\bibinfo  {journal} {Nat. Rev. Mater.}\ }\textbf {\bibinfo {volume} {3}},\ \bibinfo {pages} {460} (\bibinfo {year} {2018})}\BibitemShut {NoStop}%
\bibitem [{\citenamefont {Aieta}\ \emph {et~al.}(2012)\citenamefont {Aieta}, \citenamefont {Genevet}, \citenamefont {Kats}, \citenamefont {Yu}, \citenamefont {Blanchard}, \citenamefont {Gaburro},\ and\ \citenamefont {Capasso}}]{59PhysRevApplied.2.064002}%
  \BibitemOpen
  \bibfield  {author} {\bibinfo {author} {\bibfnamefont {F.}~\bibnamefont {Aieta}}, \bibinfo {author} {\bibfnamefont {P.}~\bibnamefont {Genevet}}, \bibinfo {author} {\bibfnamefont {M.~A.}\ \bibnamefont {Kats}}, \bibinfo {author} {\bibfnamefont {N.}~\bibnamefont {Yu}}, \bibinfo {author} {\bibfnamefont {R.}~\bibnamefont {Blanchard}}, \bibinfo {author} {\bibfnamefont {Z.}~\bibnamefont {Gaburro}},\ and\ \bibinfo {author} {\bibfnamefont {F.}~\bibnamefont {Capasso}},\ }\href@noop {} {\bibfield  {journal} {\bibinfo  {journal} {Nano Lett.}\ }\textbf {\bibinfo {volume} {12}},\ \bibinfo {pages} {4932} (\bibinfo {year} {2012})}\BibitemShut {NoStop}%
\bibitem [{\citenamefont {Liu}\ \emph {et~al.}(2018)\citenamefont {Liu}, \citenamefont {Liang}, \citenamefont {Chen},\ and\ \citenamefont {Zhu}}]{49liu2018inherent}%
  \BibitemOpen
  \bibfield  {author} {\bibinfo {author} {\bibfnamefont {T.}~\bibnamefont {Liu}}, \bibinfo {author} {\bibfnamefont {S.}~\bibnamefont {Liang}}, \bibinfo {author} {\bibfnamefont {F.}~\bibnamefont {Chen}},\ and\ \bibinfo {author} {\bibfnamefont {J.}~\bibnamefont {Zhu}},\ }\href@noop {} {\bibfield  {journal} {\bibinfo  {journal} {J. Appl. Phys.}\ }\textbf {\bibinfo {volume} {123}} (\bibinfo {year} {2018})}\BibitemShut {NoStop}%
\bibitem [{\citenamefont {Howe}()}]{s2howe2003theory}%
  \BibitemOpen
  \bibfield  {author} {\bibinfo {author} {\bibfnamefont {M.~S.}\ \bibnamefont {Howe}},\ }\href@noop {} {\emph {\bibinfo {title} {Theory of Vortex Sound}}}\ (\bibinfo  {publisher} {Cambridge University Press, 2003})\BibitemShut {NoStop}%
\bibitem [{\citenamefont {Morse}\ and\ \citenamefont {Ingard}()}]{s3morse1986theoretical}%
  \BibitemOpen
  \bibfield  {author} {\bibinfo {author} {\bibfnamefont {P.~M.}\ \bibnamefont {Morse}}\ and\ \bibinfo {author} {\bibfnamefont {K.~U.}\ \bibnamefont {Ingard}},\ }\href@noop {} {\emph {\bibinfo {title} {Theoretical Acoustics}}}\ (\bibinfo  {publisher} {Princeton University Press, 1986})\BibitemShut {NoStop}%
\bibitem [{\citenamefont {Wrobel}()}]{s4wrobel2002boundary}%
  \BibitemOpen
  \bibfield  {author} {\bibinfo {author} {\bibfnamefont {L.~C.}\ \bibnamefont {Wrobel}},\ }\href@noop {} {\emph {\bibinfo {title} {The Boundary Element Method, Volume 1: Applications in Thermo-fluids and Acoustics}}},\ Vol.~\bibinfo {volume} {1}\ (\bibinfo  {publisher} {John Wiley \& Sons, 2002})\BibitemShut {NoStop}%
\bibitem [{\citenamefont {Alaee}\ \emph {et~al.}(2018)\citenamefont {Alaee}, \citenamefont {Rockstuhl},\ and\ \citenamefont {Fernandez-Corbaton}}]{s5alaee2018electromagnetic}%
  \BibitemOpen
  \bibfield  {author} {\bibinfo {author} {\bibfnamefont {R.}~\bibnamefont {Alaee}}, \bibinfo {author} {\bibfnamefont {C.}~\bibnamefont {Rockstuhl}},\ and\ \bibinfo {author} {\bibfnamefont {I.}~\bibnamefont {Fernandez-Corbaton}},\ }\href@noop {} {\bibfield  {journal} {\bibinfo  {journal} {Opt. Commun.}\ }\textbf {\bibinfo {volume} {407}},\ \bibinfo {pages} {17} (\bibinfo {year} {2018})}\BibitemShut {NoStop}%
\bibitem [{\citenamefont {Bostr{\"o}m}(1991)}]{s6bostrom1991acoustic}%
  \BibitemOpen
  \bibfield  {author} {\bibinfo {author} {\bibfnamefont {A.}~\bibnamefont {Bostr{\"o}m}},\ }\href@noop {} {\bibfield  {journal} {\bibinfo  {journal} {J. Acoust. Soc. Am.}\ }\textbf {\bibinfo {volume} {90}},\ \bibinfo {pages} {3344} (\bibinfo {year} {1991})}\BibitemShut {NoStop}%
\end{thebibliography}%
\bibliographystyle{apsrev4-2}


\newpage


\renewcommand{\thefigure}{S\arabic{figure}}
\renewcommand{\thetable}{S\arabic{table}}
\renewcommand{\theequation}{S\arabic{equation}}
\renewcommand{\thepage}{S\arabic{page}}
\setcounter{figure}{0}
\setcounter{table}{0}
\setcounter{equation}{0}
\setcounter{page}{1} 


\pagebreak
\widetext
\begin{center}
\textbf{\large Supplementary Materials for \\
Acoustic Pancharatnam-Berry Geometric Phase}\\
Wanyue Xiao,
Wenjian Kuang, Sibo Huang, Shanjun Liang$^{\dagger}$, Din Ping Tsai, and Shubo Wang$^{\ast}$\\ 
\small$^\dagger$Corresponding author. Email: junot.liang@cpce-polyu.edu.hk\\
\small$^\ast$Corresponding author. Email: shubwang@cityu.edu.hk\\

\end{center}

\tableofcontents
\newpage


\section*{NOTE 1. Dispersion relation of the SSWs}

\noindent We consider the holey substrate shown in Fig. 1A of the main text. The reflection coefficient of the substrate can be expressed as \cite{49liu2018inherent}

\begin{equation*}
R_{m n}=\delta_{m n, 00}-\frac{2 i \tan \left(k_{0} l\right) \frac{w^{2}}{q^{2}} \xi_{00} \xi_{m n} \frac{k_{0}}{k_{z}^{(m, n)}}}{1+i \tan \left(k_{0} l\right) \frac{w^{2}}{q^{2}} \sum_{r, s=-\infty}^{+\infty} \frac{k_{0}}{k_{z}^{(r, s)}} \xi_{r s}^{2}}. \tag{S1}
\end{equation*} Here $\xi_{m n}=\operatorname{sinc}\left(k_{x}^{(m)} q / 2\right) \operatorname{sinc}\left(k_{y}^{(n)} q / 2\right)$ with $k_{x}^{(m)}=k_{x}+\frac{2 \pi m}{q}, k_{y}^{(n)}=k_{y}+\frac{2 \pi n}{q}$ and $k_{z}^{(m, n)}=\sqrt{k_{0}^{2}-\left(k_{x}^{(m)}\right)^{2}-\left(k_{y}^{(n)}\right)^{2}}$, where $k_{x}$ and $k_{y}$ are the $x$ and $y$ components of the incident wave vector, respectively, and $m, n, r$, and $s$ are integers. The dispersion relation of the SSWs can be obtained by analyzing the poles of the reflection coefficient in Eq. (S1) \cite{49liu2018inherent}

\begin{equation*}
1-k_{0} \tan \left(k_{0} l\right) \frac{w^{2}}{q^{2}} \sum_{m, n=-\infty}^{+\infty} \frac{\xi_{m n}^{2}}{\sqrt{\left(\beta^{(m, n)}\right)^{2}-k_{0}^{2}}}=0. \tag{S2}
\end{equation*}Here $\beta^{(m, n)}=\sqrt{\left(k_{x}^{(m)}\right)^{2}+\left(k_{y}^{(n)}\right)^{2}}$ is the propagation constant of the SSWs. For the SSW propagating in $x$ direction with $m=n=0$, we can obtain

\begin{equation*}
1-\tan \left(k_{0} l\right) \frac{w^{2}}{q^{2}} \frac{\operatorname{sinc}^{2}\left(\frac{\beta a}{2}\right)}{\sqrt{\frac{\beta^{2}}{k_{0}^{2}}-1}}=0, \tag{S3}
\end{equation*}where $\beta=\beta^{(0,0)}=k_{x}^{(0)}$. In the deep subwavelength limit $q \ll \lambda$, $\operatorname{sinc}(\beta a / 2) \approx 1$, \textcolor{black}{Eq. (S3) is reduced to an isotropic dispersion relation}

\begin{equation*}
\beta=k_{0} \sqrt{1+\left(\frac{w}{q}\right)^{4} \tan ^{2}\left(k_{0} l\right)}. \tag{S4}
\end{equation*}This is the dispersion relation shown in Fig. 1B of the main text. 
\textcolor{black}{Figure S1 shows the simulated isofrequency contours of the dispersion
relation. Clearly, the holey substrate is effectively homogeneous and isotropic for the SSWs in the considered frequency range.}
\begin{figure}[htp]
\centering
\includegraphics{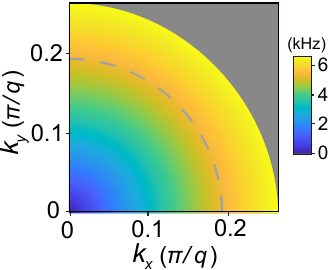}
\caption{The isofrequency contour of the dispersion relation in the $k_x-k_y$ plane. The gray dash line denotes the working frequency of the holey substrate.
}\label{Fig. S1}
\end{figure}

\section*{NOTE 2. Multipole expansion of the meta-atom scattering field}
\noindent The linear acoustic wave equation with generic sources can be written as \cite{s2howe2003theory}

\begin{gather*}
\frac{1}{c^{2} \rho_{0}} \frac{\partial p(\mathbf{r}, t)}{\partial t}+\nabla \cdot \mathbf{v}(\mathbf{r}, t)=q(\mathbf{r}, t),  \tag{S5}\\
\rho_{0} \frac{\partial \mathbf{v}(\mathbf{r}, t)}{\partial t}+\nabla p(\mathbf{r}, t)=\mathbf{F}(\mathbf{r}, t), \tag{S6}
\end{gather*}where $q(\mathbf{r}, t)$ is volume change rate which can be related to monopole density by $M(\mathbf{r}, t)=$ $\rho_{0} \frac{\partial q(\mathbf{r}, t)}{\partial t}, \mathbf{F}(\mathbf{r}, t)$ is the force density which can be defined as the dipole density $\mathbf{D}(\mathbf{r}, t)=\mathbf{F}(\mathbf{r}, t)$. Equations (S5) and (S6) correspond to the conservation law of mass and the conservation law of momentum, respectively. Based on the two equations, we can obtain the inhomogeneous Helmholtz wave equation for the monochromatic time-harmonic acoustic wave of frequency $\omega$:

\begin{equation*}
\left(\nabla^{2}+k_{0}^{2}\right) p(\mathbf{r})=\nabla \cdot \mathbf{D}(\mathbf{r})-M(\mathbf{r}). \tag{S7}
\end{equation*}The corresponding Green's function $G\left(\mathbf{r}, \mathbf{r}^{\prime}\right)$ is

\begin{equation*}
\left(\nabla^{2}+k_{0}^{2}\right) G\left(\mathbf{r}, \mathbf{r}^{\prime}\right)=-\delta\left(\mathbf{r}-\mathbf{r}^{\prime}\right), \tag{S8}
\end{equation*}and the general solution satisfying radiation boundary condition is $G\left(\mathbf{r}, \mathbf{r}^{\prime}\right)=\frac{e^{i k_{0}\left|\mathbf{r}-\mathbf{r}^{\prime}\right|}}{4 \pi|\mathbf{r}-\mathbf{r}{\prime}|}$. The pressure field outside the source volume $\tau$ can be obtained by using the Green's function \cite{s3morse1986theoretical}

\begin{equation*}
p(\mathbf{r})=-\int_{\tau} \nabla \cdot \mathbf{D}\left(\mathbf{r}^{\prime}\right) G\left(\mathbf{r}, \mathbf{r}^{\prime}\right) d^{3} r^{\prime}+\int_{\tau} M\left(\mathbf{r}^{\prime}\right) G\left(\mathbf{r}, \mathbf{r}^{\prime}\right) d^{3} r^{\prime} \tag{S9}
\end{equation*}
$$
=\int_{\tau} \mathbf{D}\left(\mathbf{r}^{\prime}\right) \cdot \nabla^{\prime} G\left(\mathbf{r}, \mathbf{r}^{\prime}\right) d^{3} r^{\prime}+\int_{\tau} M\left(\mathbf{r}^{\prime}\right) G\left(\mathbf{r}, \mathbf{r}^{\prime}\right) d^{3} r^{\prime}.
$$If the sources only distribute on the boundary of the scatterer, the above equation reduces to

\begin{equation*}
p(\mathbf{r})=\int_{\partial \tau} \mathbf{D}\left(\mathbf{r}^{\prime}\right) \cdot \nabla^{\prime} G\left(\mathbf{r}, \mathbf{r}^{\prime}\right) d^{2} r^{\prime}+\int_{\partial \tau} M\left(\mathbf{r}^{\prime}\right) G\left(\mathbf{r}, \mathbf{r}^{\prime}\right) d^{2} r^{\prime}, \tag{$\mathrm{S} 9^{\prime}$}
\end{equation*}where $\mathbf{D}\left(\mathbf{r}^{\prime}\right)$ and $M\left(\mathbf{r}^{\prime}\right)$ now represent surface densities of the sources. According to the boundary element method \cite{s4wrobel2002boundary}, the pressure field can be expressed as 

\begin{equation*}
p(\mathbf{r})=\int_{\partial \tau}\left[p\left(\mathbf{r}^{\prime}\right) \mathbf{n} \cdot \nabla^{\prime} G\left(\mathbf{r}, \mathbf{r}^{\prime}\right)-G\left(\mathbf{r}, \mathbf{r}^{\prime}\right) \mathbf{n} \cdot \nabla^{\prime} p\left(\mathbf{r}^{\prime}\right)\right] d^{2} r^{\prime}, \tag{S10}
\end{equation*}where $\mathbf{n}$ is the unit normal vector on the boundary $\partial \tau;$ $p\left(\mathbf{r}^{\prime}\right) \mathbf{n}$ and $-\mathbf{n} \cdot \nabla^{\prime} p\left(\mathbf{r}^{\prime}\right)$ are the boundary source densities. Compare Eq. (S10) with Eq. ($\mathrm{S} 9^{\prime}$), we find that $p\left(\mathbf{r}^{\prime}\right) \mathbf{n}$ and $-\mathbf{n} \cdot \nabla^{\prime} p\left(\mathbf{r}^{\prime}\right)$ correspond to the monopole source density and dipole source density, respectively. For a Neumann-type boundary condition $\mathbf{n} \cdot \nabla p=0$ (corresponding to a hard boundary), the normal component of the velocity field is zero $v_{n} \propto \mathbf{n} \cdot \nabla p=0$, and only the dipole source density $\mathbf{D}\left(\mathbf{r}^{\prime}\right)=p\left(\mathbf{r}^{\prime}\right) \mathbf{n}$ exists on the boundary. For a Dirichlet-type boundary condition $p=0$ (corresponding to a soft boundary), the pressure field is zero, and only the monopole source $M\left(\mathbf{r}^{\prime}\right)=-\mathbf{n} \cdot \nabla^{\prime} p\left(\mathbf{r}^{\prime}\right)$ exists on the boundary.

For a rigid scatterer under the incidence of external sound waves, as shown in Fig. S2, only dipole density can be induced on the surface with  $\mathbf{D}=\mathbf{n} p_{t}$, where $p_{t}$ is surface total pressure. Thus, the pressure field in Eq. ($\mathrm{S} 9^{\prime}$) reduces to

\begin{figure}[t]
\centering
\includegraphics[width=0.3\linewidth]{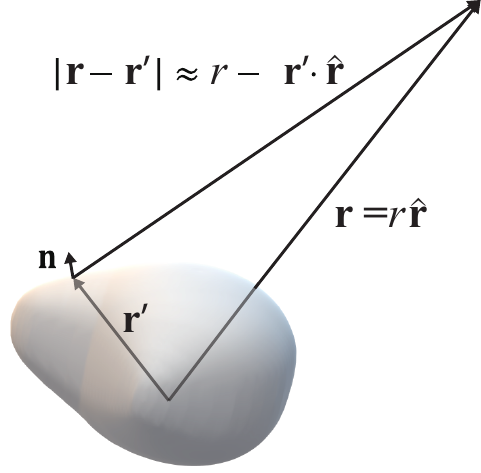}
\caption{Illustration of multipolar expansion. Schematic view of the rigid acoustic scatter with arbitrary shaped boundary.}\label{Fig. S2}
\end{figure}

\begin{equation*}
p(\mathbf{r})=\int_{\partial \tau} \mathbf{D}\left(\mathbf{r}^{\prime}\right) \cdot \nabla^{\prime} G\left(\mathbf{r}, \mathbf{r}^{\prime}\right) d^{2} r^{\prime}, \tag{S11}
\end{equation*}Equivalently,

\begin{equation*}
p(\mathbf{r})=\int_{\partial \tau} \mathbf{D}\left(\mathbf{r}^{\prime}\right) \cdot\left[\frac{\mathbf{R}}{R}\left(-i k_{0}+\frac{1}{R}\right) G\left(\mathbf{r}, \mathbf{r}^{\prime}\right)\right] d^{2} r^{\prime}, \tag{S12}
\end{equation*}where $R=\left|\mathbf{r}-\mathbf{r}^{\prime}\right|$. Using the approximation $\left|\mathbf{r}-\mathbf{r}^{\prime}\right| \approx r-\mathbf{r}^{\prime} \cdot \hat{\mathbf{r}}$ with $\hat{\mathbf{r}}=\mathbf{r} / r$, the scattering far field of a subwavelength scatterer can be expanded as

\begin{equation*}
p(\mathbf{r})=\frac{-i k_{0}}{4 \pi} \frac{e^{\mathrm{i} k_{0} r}}{r} \sum_{n} \frac{\left(-\mathrm{i} k_{0}\right)^{n}}{n!} \int_{\partial \tau}\left[\mathbf{D}\left(\mathbf{r}^{\prime}\right) \cdot \hat{\mathbf{r}}\right]\left(\mathbf{r}^{\prime} \cdot \hat{\mathbf{r}}\right)^{n} d^{2} r^{\prime}, \tag{S13}
\end{equation*}The contribution from the first term is

\begin{equation*}
p(\mathbf{r})=\frac{-i k_{0}}{4 \pi} \frac{e^{\mathbf{i} k_{0} r}}{r} \hat{\mathbf{r}} \cdot \int_{\partial \tau} \mathbf{D}\left(\mathbf{r}^{\prime}\right) d^{2} r^{\prime}, \tag{S14}
\end{equation*}which corresponds to the far field of an acoustic dipole \cite{s3morse1986theoretical}

\begin{equation*}
p(\mathbf{r})=\frac{-i k_{0}}{4 \pi} \frac{e^{\mathrm{i} k_{0} r}}{r} \hat{\mathbf{r}} \cdot \mathbf{d}, \tag{S15}
\end{equation*}with the dipole moment

\begin{equation*}
\mathbf{d}=\int_{\partial \tau} \mathbf{D}\left(\mathbf{r}^{\prime}\right) d^{2} r^{\prime}. \tag{S16}
\end{equation*}The contribution from the second term in Eq. (S13) is

\begin{equation*}
p(\mathbf{r})=\frac{-i k_{0}}{4 \pi} \frac{e^{i k_{0} r}}{r}\left(-i k_{0}\right) \int_{\partial \tau}\left[\mathbf{D}\left(\mathbf{r}^{\prime}\right) \cdot \hat{\mathbf{r}}\right]\left(\mathbf{r}^{\prime} \cdot \hat{\mathbf{r}}\right) d^{2} r^{\prime}, \tag{S17}
\end{equation*}which can be rewritten as

\begin{equation*}
p(\mathbf{r})=\frac{-k_{0}^{2}}{4 \pi} \frac{e^{\mathrm{i} k_{0} r}}{r} \hat{\mathbf{r}}\left\{\int_{\partial \tau} \mathbf{D}\left(\mathbf{r}^{\prime}\right) \otimes \mathbf{r}^{\prime} d^{2} r^{\prime}\right\} \hat{\mathbf{r}}^{\mathrm{T}}. \tag{S18}
\end{equation*}Denote the tensor inside the bracket as $\overline{\mathbf{M}}$:

\begin{equation*}
\overline{\mathbf{M}}=\int_{\partial \tau} \mathbf{D}\left(\mathbf{r}^{\prime}\right) \otimes \mathbf{r}^{\prime} d^{2} r^{\prime}=\left(\begin{array}{lll}
M_{x x} & M_{x y} & M_{x z} \\
M_{y x} & M_{y y} & M_{y z} \\
M_{z x} & M_{z y} & M_{z z} 
\end{array}\right). \tag{S19}
\
\end{equation*}For $\overline{\mathbf{M}}$ with elements $M_{x x}=M_{y y}=M_{z z} \neq 0$ and other elements being zero, Eq. (S18) is reduced to

\begin{equation*}
p(\mathbf{r})=\frac{-k_{0}^{2}}{4 \pi} \frac{e^{\mathrm{i} k_{0} r}}{r} \frac{1}{3} \int_{\partial \tau} \mathbf{D}\left(\mathbf{r}^{\prime}\right) \cdot \mathbf{r}^{\prime} d^{2} r^{\prime}, \tag{S20}
\end{equation*}which takes the form of an acoustic monopole \cite{s3morse1986theoretical}

\begin{equation*}
p(\mathbf{r})=m \frac{e^{\mathrm{i} k_{0} r}}{4 \pi r} \tag{S21}
\end{equation*}with amplitude

\begin{equation*}
m=\frac{-k_{0}^{2}}{3} \int_{\partial \tau} \mathbf{D}\left(\mathbf{r}^{\prime}\right) \cdot \mathbf{r}^{\prime} d^{2} r^{\prime}. \tag{S22}
\end{equation*}For $\overline{\mathbf{M}}$ with other matrix values, it corresponds to an acoustic quadrupole, which usually has much smaller contribution compared to that of monopole and dipole in the deep subwavelength regime.

The expressions of monopole and dipole given by Eq. (S16) and (S22) only apply to the scatterers with geometric dimensions much smaller than the wavelength (i.e., in the deep subwavelength regime). For large scatterers or high frequencies, correction factors have to be introduced to obtain accurate multipoles \cite{s5alaee2018electromagnetic}, in which case the monopole and dipole can be determined as

\begin{gather*}
m=\frac{-k_{0}^{2}}{3} \int_{\partial \tau} \mathbf{D}\left(\mathbf{r}^{\prime}\right) \cdot \mathbf{r}^{\prime} \frac{3 j_{1}\left(k_{0} r^{\prime}\right)}{k_{0} r^{\prime}} d^{2} r^{\prime},  \tag{S23}\\\\
\mathbf{d}=\int_{\partial \tau} \mathbf{D}\left(\mathbf{r}^{\prime}\right) \frac{3 j_{1}\left(k_{0} r^{\prime}\right)}{k_{0} r^{\prime}} d^{2} r^{\prime}, \tag{S24}
\end{gather*}where $j_{1}\left(k_{0} r^{\prime}\right)$ is the spherical Bessel function.

\section*{NOTE 3. Scattering cross sections of the multipoles}

\noindent The pressure far field of a monopole is given by Eq. (S21). The corresponding velocity far field in radial direction is

\begin{equation*}
v_{r}=\frac{m}{\rho_{0} c} \frac{e^{i k_{0} r}}{4 \pi r}. \tag{S25}
\end{equation*}The time-averaged intensity in the far field reads

\begin{equation*}
I_{r}=\frac{1}{2} \operatorname{Re}\left(p v_{r}^{*}\right)=\frac{m^{2}}{32 \pi^{2} r^{2} \rho_{0} c}. \tag{S26}
\end{equation*}The radiation power is

\begin{equation*}
P=\oint I_{r} r^{2} d \Omega=\frac{m^{2}}{8 \pi \rho_{0} c}. \tag{S27}
\end{equation*}Thus, the scattering cross section due to monopole is

\begin{equation*}
C_{\mathrm{sca}}^{m}=\frac{P}{I_{0}}=\frac{m^{2}}{8 \pi \rho_{0} c I_{0}}, \tag{S28}
\end{equation*}where $I_{0}$ is the intensity of the incident wave.

For an acoustic dipole, the pressure far field is given by Eq. (S15). The velocity far field can be written as

\begin{equation*}
v_{r}=\frac{-i k_{0}}{\rho_{0} c} \frac{e^{i k_{0} r}}{4 \pi r} \hat{\mathbf{r}} \cdot \mathbf{d}=\frac{-i k_{0}}{\rho_{0} c} \frac{e^{i k_{0} r}}{4 \pi r}\left(d_{x} \sin \theta \cos \varphi+d_{y} \sin \theta \sin \varphi+d_{z} \cos \varphi\right). \tag{S29}
\end{equation*}The scattering cross section due to the dipole can be obtained as in the monopole case:

\begin{equation*}
C_{\mathrm{sca}}^{\mathbf{d}}=\frac{k_{0}^{2}}{24 \pi \rho_{0} c I_{0}}|\mathbf{d}|^{2}. \tag{S30}
\end{equation*}

\section*{NOTE 4. Scattering properties of the thin plate}

\noindent We will show that the rigid thin plate can be safely used as a meta-atom to support \textcolor{black}{dominant} dipole mode as long as its thickness $(t)$ is much smaller than the side lengths $(a, b)$ and the side length is not too large compared with the wavelength. We first consider a rigid plate with negligible thickness $(t \rightarrow 0$) and discuss its scattered far field as a function of side lengths $a$ and $b$, then we explore the influence of the thickness $t$.

\begin{figure}[t]
\centering
\includegraphics[width=0.95\linewidth]{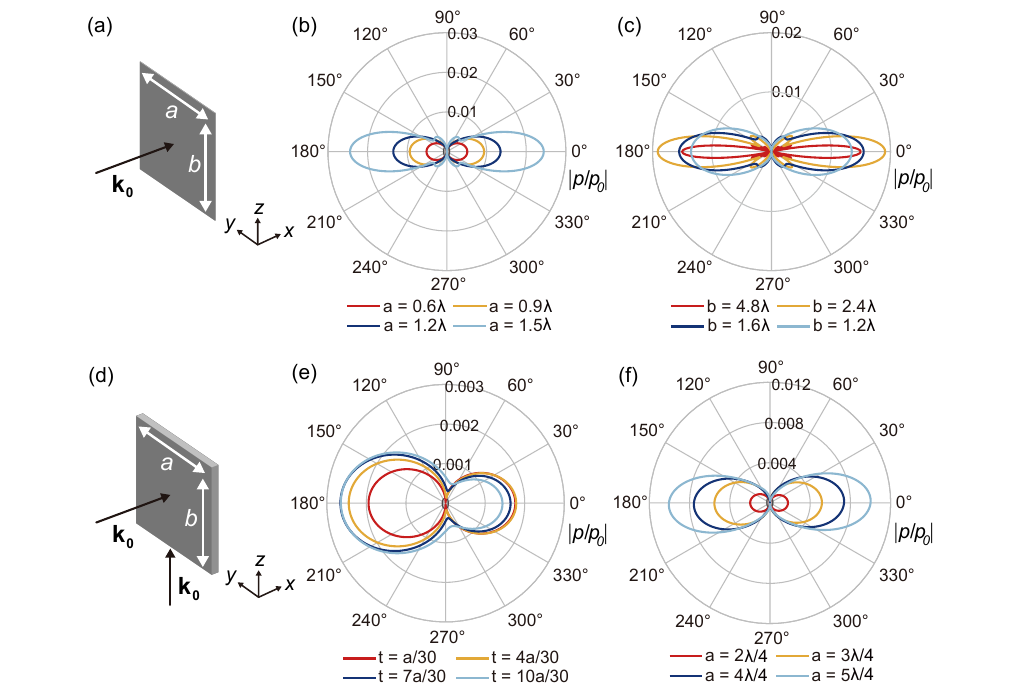} 
\caption{A rigid thin plate with negligible thickness and side lengths $a$ and $b$. (b) Scattered pressure amplitude in the far field for the plate with different side length $a$ (assuming $a=b$). (c) Scattered pressure amplitude in the far field for the plate with the same area $a \times b=(1.2 \lambda)^{2}$ but different side length $b$ (or $a$). (d) A rigid thin plate with finite thickness $t$ and side lengths $a$ and $b$. Scattered pressure amplitude in the far field for the plate (e) with different thickness (assuming $a=$ $b=\lambda / 2$) and (f) with different side length $a$ (assuming $t=a / 30$).}
\end{figure}

Figure S3(a) shows a rigid plate with a negligible thickness. We assume $a=b$, and the area of the plate is $a^{2}$. Under the incidence of a plane wave propagating along $+x$ direction, the scattered pressure amplitude (normalized by the incident pressure amplitude $p_{0}$) in the far field (at a distance of $100 \lambda$ away from the plate's center and on the $x z$-mirror plane) is obtained for different area $a^{2}$ based on the analytical expression given in Ref. \cite{s6bostrom1991acoustic}. The results are shown in Fig. S3(b). Generally, for a subwavelength plate, its scattered field pattern corresponds to a \textcolor{black}{dominant} dipole mode. Then, we fix the area of the plate to be $a \times b=(1.2 \lambda)^{2}$ and change $b$ from $1.2 \lambda$ to $4.8 \lambda$. The calculated farfield pressure amplitude on the $x z$-mirror plane is shown in Fig. S3(c). Obviously, higher order modes will be excited if $b$ is much larger than the wavelength. Thus, a rigid plate with negligible thickness can support \textcolor{black}{dominant} dipole mode as long as its side lengths satisfy $a<\lambda$ and $b<\lambda$.

Figure S3(d) shows a rigid plate with a finite thickness $t$ and side lengths $a=b=\lambda / 2$. Under the incidence of a circularly polarized velocity field $\mathbf{v}=\left(v_{x}, 0, v_{z}\right)=\frac{p_{0}}{\sqrt{2} \rho_{0} c}\left(e^{i k_{0} x}, 0, i e^{i k_{0} z}\right)$, we numerically simulate the scattered pressure amplitude in the far field (on $x z$-mirror plane) for different thickness $t$. The results are shown in Fig. S3(e). As seen, the scattered field pattern shows a dipole mode when $t \ll a$. As $t$ increases, asymmetric scattered field pattern will appear due to the interference of dipole and monopole. The results in Fig. S3(e) indicate that $t=a / 30$ is a safe value for exciting \textcolor{black}{dominant} dipole mode. Figure S3(f) shows the field pattern for different side length $a$ and $t=a / 30$, confirming \textcolor{black}{dominant} dipole mode is indeed excited.

To summarize, a general rigid thin plate can be used as a meta-atom that support \textcolor{black}{dominant} dipole mode as long as $t \ll a<\lambda$ and $t \ll b<\lambda$.

\section*{NOTE 5. Effect of monopole on the PB geometric phase}

\begin{figure}[t!]
\centering
\includegraphics{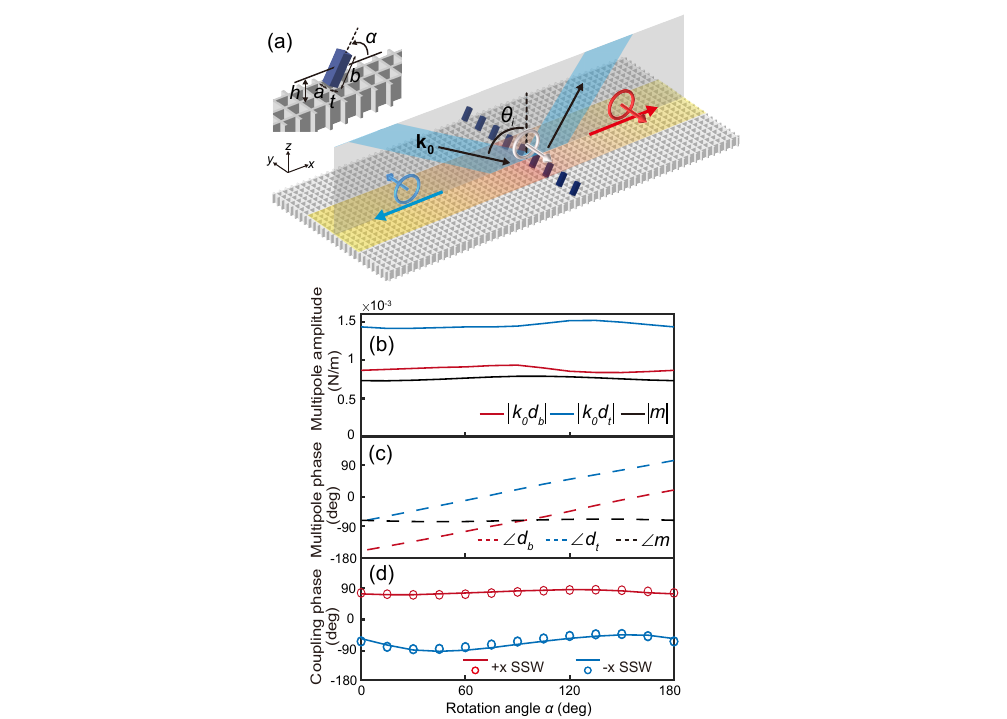} 
\caption{(a) Schematic of the metasurface. The inset shows the meta-atom with dimensions $a=$ $5.3 \mathrm{~mm}$ $(\sim \lambda / 12), b=10.7 \mathrm{~mm}$ $(\sim \lambda / 6), t=5.3 \mathrm{~mm}$ $(\sim \lambda / 12)$ and $h=6.5 \mathrm{~mm}$. (b) Amplitude and (c) phase of the induced acoustic multipoles $d_{b}, d_{t}$ and $m$ in the meta-atom. (d) The coupling phases of the SSWs propagating in $+x$ and $-x$ direction as a function of the rotation angle. The solid lines are simulation results, and the symbols denote the analytical results obtained with coupled mode theory.}
\end{figure}

\noindent We study the effect of monopole induced in the meta-atom on the PB geometric phase. Let us consider a meta-atom plate with dimensions $a \times b \times t=5.3 \mathrm{~mm} \times 10.7 \mathrm{~mm} \times 5.3 \mathrm{~mm}$, as shown in the inset of Fig. S4(a), which will induce monopole in addition to dipole according to the above discussions. We arrange the meta-atoms on the substrate (at $h=6.5 \mathrm{~mm}$) periodically along the $y$ direction with a period of $p=16.5\mathrm{~mm}$ $(\sim \lambda / 4)$. The incident acoustic plane wave propagate in the $x z$-plane with the incident angle $\theta_{i}=73^{\circ}$ and frequency $f=5378 \mathrm{~Hz}$. Under the excitation of the background circularly polarized field (i.e., the total field due to the interference of the incident and reflected waves), the meta-atom gives rise to two dipole components $d_{t}$ (in the direction along the side $t$) and $d_{b}$ (in the direction along the side $b$), and a monopole moment $m$. Figures S4(b) and S4(c) show the amplitude and phase of the excited multipoles as a function of the rotation angle $\alpha$ of the meta-atom, respectively, calculated by using Eqs. (S16) and (S22). We see that the dipole is generally elliptically polarized, and its direction rotates with the meta-atom. Figure S4(d) shows the simulated phases of the SSWs propagating in $+x$ and $-x$ directions. Clearly, the phases cannot cover $2 \pi$, which is different from the cases discussed in the main text. This is attributed to the monopole excited in the meta-atom, which can be understood with a coupled mode theory \cite{64long2020symmetry}.

The eigen fields of the SSWs can be expressed as

\begin{equation*}
|\mathbf{F}\rangle=(v_x, v_z, p)^\mathrm{T}= (\pm n_{\mathrm{eff}}, +i \gamma, -i)^\mathrm{T}, \tag{S31}
\end{equation*}where `T’ denotes transpose, `$ + $' (`$ - $') is for the SSW propagating in the $+x$ $(-x)$ direction. The source multipoles induced in the meta-atom can be expressed as

\begin{equation*}
|\mathbf{S}\rangle=(k_0 d_x, k_0 d_z, m)^\mathrm{T}. \tag{S32}
\end{equation*}The coupling coefficient between the multipoles and the SSWs can be determined as

\begin{equation*}
C=\langle\mathbf{F}| \mathbf{S}\rangle.  \tag{S33}
\end{equation*} Using the numerically determined eigen fields of the SSWs and the multipole moments in Figs. S4(b) and S4(c), we determined the coupling phase arg($C$) with Eq. (S33). The results are shown in Fig. S4(d) by the circles, which have good consistency with the simulation results and demonstrating the validity of the coupled mode theory.

We can apply the coupled mode theory to understand the contribution of the monopole to the phase of the SSWs. We decompose Eq. (S33) into a monopole term and dipole term

\begin{equation*}
C=\langle\mathbf{F}|\mathbf{S}\rangle=\left\langle\mathbf{F}_{\mathbf{v}}|\mathbf{S}_{\mathbf{v}}\right\rangle+i m, \tag{S34}
\end{equation*}where $\left|\mathbf{F}_{\mathbf{v}}\right\rangle= (\pm n_{\mathrm{eff}},+i \gamma)^\mathrm{T}$ and $\left|\mathbf{S}_{\mathbf{v}}\right\rangle=(k_0 d_x, k_0 d_z)^\mathrm{T}=\widehat{\mathbb{R}}(\alpha)\left|\mathbf{S}_{\mathbf{v}}^{\prime}\right\rangle$ with $\widehat{\mathbb{R}}(\alpha)=$ $\left(\begin{array}{cc}\cos \alpha & -\sin \alpha \\ \sin \alpha & \cos \alpha\end{array}\right)$ being the rotation matrix and $\left|\mathbf{S}_{\mathbf{v}}^{\prime}\right\rangle=(k_0 d_b,k_0 d_t)^\mathrm{T}$ being the dipole in the local frame of the meta-atom. The dipole is induced by the background fields and can be expressed as:

\begin{equation*}
\left|\mathbf{S}_{\mathbf{v}}^{\prime}\right\rangle=\left\{\left\langle\mathbf{S}_{\mathbf{v}}^{\mathrm{e}}\right| \widehat{\mathbb{R}}(-\alpha)\left|\mathbf{F}_{\mathbf{v}}^0\right\rangle\right\}\left|\mathbf{S}_{\mathbf{v}}^{\mathrm{e}}\right\rangle, \tag{S35}
\end{equation*}where $\left|\mathbf{S}_{\mathbf{v}}^\mathrm{e}\right\rangle$ is the eigen dipole mode of the meta-atom with rotation angle $\alpha=0$ and $\left|\mathbf{F}_{\mathbf{v}}^0\right\rangle$ is the background velocity field (i.e., the interference field). Substituting it into Eq. (S34), we obtain

\begin{align*}
\begin{aligned}
& C=\left\langle\mathbf{F}_{\mathbf{v}}\right| \widehat{\mathbb{R}}(\alpha)\left\{\left\langle\mathbf{S}_{\mathbf{v}}^{\mathrm{e}}\right| \widehat{\mathbb{R}}(-\alpha)\left|\mathbf{F}_{\mathbf{v}}^0\right\rangle\right\}\left|\mathbf{S}_{\mathbf{v}}^{\mathrm{e}}\right\rangle+i m \\
& =\textcolor{blue}{\left\{\left\langle\mathbf{F}_{\mathbf{v}}\right| \widehat{\mathbb{R}}(\alpha)\left|\mathbf{S}_{\mathbf{v}}^{\mathrm{e}}\right\rangle\right\}}\textcolor{red}{\left\{\left\langle\mathbf{S}_{\mathbf{v}}^{\mathrm{e}}\right| \widehat{\mathbb{R}}(-\alpha)\left|\mathbf{F}_{\mathbf{v}}^0\right\rangle\right\}}+i m .
\end{aligned} \tag{S36}
\end{align*}Here, the red part characterizes the coupling between the background field and the meta-atom, and the blue part characterizes the coupling between the meta-atom and the SSWs. Apparently, the rotation of the meta-atom only affects the dipole term in the coupling coefficient, it does not affect the monopole term due to the isotropy of the monopole fields. In other words, the PB geometric phase can only manifest in the dipole term. The monopole term will interfere with the dipole term and contribute to the phase of the total field, which makes the PB phase ambiguous.

\section*{NOTE 6. Optimization for the amplitude of the SSWs}

\noindent{In the main text, we demonstrate the physics of acoustic PB phase by using probably the simplest meta-atom structure (i.e., a thin plate). This structure leads to nonuniform amplitude of the modulated waves due to the different coupling strength between the SSWs and the meta-atoms with different rotation angles. Here, we propose two approaches to improve the amplitude uniformity and strength. }\\
\noindent{\textbf{Method 1: Meta-atom thin plates with tailored dimensions}}\\
In the main text, we construct the metasurfaces using identical thin plates (i.e., meta-atoms). To improve the uniformity of the SSW’s amplitude, we can tailor the side length $b$ of the plates (fixing other parameters) to control the coupling strength. Figure S5(a) shows the resulting SSW’s amplitude and phase at different rotation angles $\alpha$, and the corresponding values of $b$ are shown in Fig. S5(b). We see that the phase can cover $2\pi$ and the amplitude is approximately uniform over all the rotation angles. Thus, this approach can improve the uniformity of the modulated waves.

\begin{figure}[htp!]
\centering
\includegraphics{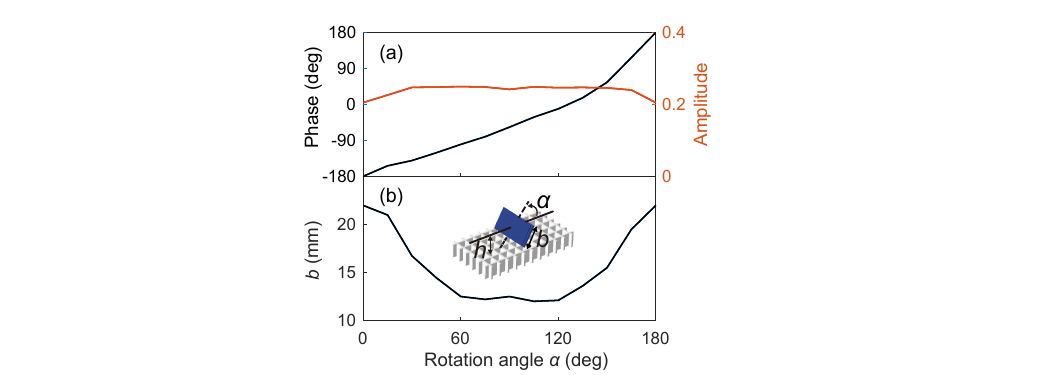} 
\caption{(a) Phase and amplitude of the SSW (propagating in $+x$ direction) for the meta-atoms with different rotation angles $\alpha$. (b) The side lengths $b$ of the meta-atom corresponding to (a).}
\end{figure}

\noindent\textcolor{black}{{\textbf{Method 2: Resonant meta-atoms and substrate supporting SSWs with a large effective index}}\\
The modulated waves' amplitude in Fig. 3B is not uniform because the SSWs in the main text are elliptically polarized. This leads to the different coupling strength between the SSWs and the meta-atoms with different rotation angles, and thus nonuniform amplitude of the modulated wave. Therefore, another method to achieve uniform amplitude is to use the substrate that supports circularly polarized SSWs ($\left|S_{3}\right|=\frac{2 n_{\text {eff }} \sqrt{n_{\text {eff }}^{2}-1}}{2 n_{\text {eff }}^{2}-1}$ denotes the ellipticity of the SSWs and $\left|S_{3}\right|=1$ for circularly polarized SSWs). Such SSWs (with a larger effective index) exhibit better confinement and stronger coupling with the meta-atoms, contributing to the enhancement of the amplitude. The SSWs' amplitude can be further enhanced if we exploit resonant meta-atoms. Based on these considerations, we propose another design of the acoustic PB metasurface to achieve strong and uniform amplitude of the modulated SSWs.}

\begin{figure}[t!]
\centering
\includegraphics{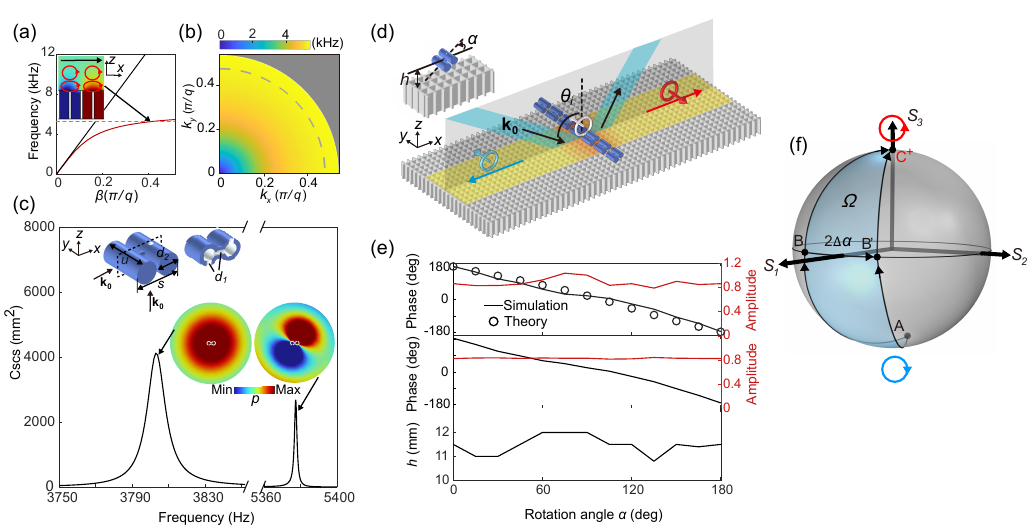} 
\caption{(a) Dispersion relation of the SSWs supported by the holey substrate with hole depth $l=13.5 \mathrm{~mm}$. The inset shows the pressure field and the polarization of the velocity field for the SSW propagating in $+x$ direction at $f=5378 \mathrm{~Hz}$. (b) The isofrequency contour of the dispersion relation in the $k_{x}-k_{y}$ plane. The gray dash line marks the working frequency of the holey substrate. (c) Scattering cross section of the meta-atom as a function of frequency. The insets show the meta-atom and its cross-sectional view. The parameters are $u=11 \mathrm{~mm}, s=10.5$ $\mathrm{mm}, d_{1}=0.8 \mathrm{~mm}$ (diameter of the openings and the tube connecting the two cavities), and $d_{2}=5 \mathrm{~mm}$. The wall thickness is 1 mm. (d) Acoustic PB metasurface under the incidence of a plane wave. The incident angle is $\theta_{i}=55^{\circ}$. The inset shows the meta-atom with $h=11 \mathrm{~mm}$ and period $p=16.5 \mathrm{~mm}$. (e) Upper panel: the PB phase and amplitude of +SSW for the meta-atoms with different rotation angles $\alpha$. The solid lines denote the simulation results. The circles denote the analytical result obtained by evaluating the solid angle in the Poincaré sphere. Middle panel: the phase and amplitude of +SSW after adjusting $h$. Lower panel: the height $h$ of the metaatoms corresponding to the middle panel. (f) Evolution trajectories of velocity polarization on the Poincaré sphere for +SSW.}
\end{figure}

We use a holey substrate similar to the one in the main text but with larger depth $l$ of the holes. The dispersion relation of the SSWs is shown as the red solid line in Fig. S6(a). The black solid line denotes the sound dispersion in free space. The inset shows the pressure field of the SSW propagating in $+x$ direction at $f=5378 \mathrm{~Hz}$, which has an effective index $n_{\text {eff }}=2.7$ (the SSWs in the main text have $n_{\text {eff }}=1.1$). The polarization of the velocity field generally changes with $h$ (the height above the substrate). At a large $h$, the polarization approaches $\left|S_{3}\right|=1$. The red arrows in the inset in Fig. S6(a) indicate the polarization of the velocity field at $h=2 \mathrm{~mm}$ and 8 mm, respectively. The velocity field of this SSW is approximately circularly polarized at $h>8 \mathrm{~mm}$. Figure S6(b) shows the simulated isofrequency contour of the dispersion relation, indicating the substrate is effectively homogeneous and isotropic for the SSWs in the considered frequency range.

We then design a subwavelength meta-atom supporting acoustic dipole resonance. The meta-atom comprises two Helmholtz cavities connected by tubes, as shown in the inset of Fig. S6(c) (including a cross-sectional view of the meta-atom). The scattering crosssection of the meta-atom under the incidence of a velocity field $\mathbf{v}=\left(v_{x}, 0, v_{z}\right)=\frac{p_{0}}{\sqrt{2} \rho_{0} c}\left(e^{i k_{0} x}, 0, e^{i\left(k_{0} z-\frac{\pi}{2}\right)}\right)$ is shown in Fig. S6(c) as the solid black line. The first resonance peak at 3803 Hz corresponds to the isotropic monopole mode. The second peak at 5378 Hz corresponds to the dipole resonance.

\textcolor{black}{
Using the above holey substrate and resonant meta-atoms, we construct a metasurface shown in Fig. S6(d). The meta-atoms locate at $h=11 \mathrm{~mm}$ above the substrate. We consider a plane wave illuminating the holey substrate with incident angle $\theta_{i}=55^{\circ}$. The white, blue, and red arrows denote the spin of the background velocity field ($S_{3}=-0.92$ ), the SSW propagating in $-x$ direction ($S_{3}=-1$), and the SSW propagating in $+x$ direction $\left(S_{3}=1\right)$, respectively. We note that the spin direction of the background field is opposite to that in the main text. This is because the polarization of the background velocity field varies with $h$ (as we discussed in Fig. 1D). Here, the meta-atoms locate at a larger $h$ where the spin direction is reversed. In this case, the SSW propagating in $+x$ direction (+SSW) can carry the PB phase covering $2 \pi$ due to the spin flipping. The numerically simulated PB phase for the meta-atoms with different rotation angles $\alpha$ are shown in the upper panel of Fig. S6(e) by the black solid line. We also analytically calculated the PB phase by evaluating the solid angle $\Omega$ in the Poincaré sphere, as shown in Fig. S6(f), where A denotes the polarization of the background velocity field; $\mathrm{C}^{+}$denotes the polarization of +SSW; B and $\mathrm{B}^{\prime}$ denote the polarization of the meta-atoms with different rotation angles. The analytical result is denoted by the black circles in the upper panel of Fig. S6(e), showing good consistency with the numerical result. In addition, we show the modulated SSW amplitude (red solid line) in the upper panel of Fig. S6(e). It is noticed that the amplitude is much stronger (0.83) than that in the main text and exhibits good uniformity at different rotation angles. The uniformity of the amplitude can be further improved by adjusting the meta-atoms' height $h$, as shown in the middle panel of Fig. S6(e). The corresponding $h$ values of the meta-atoms are shown in the bottom panel of Fig. S6(e). Therefore, this new design of acoustic PB metasurface can realize the desired modulations of the SSWs with \emph{both uniform and strong amplitude}.
}

\end{document}